\documentclass[12pt, draftclsnofoot, onecolumn]{IEEEtran}

\usepackage{graphicx}
\DeclareGraphicsExtensions{.pdf,.jpeg,.png}
\usepackage{epstopdf}
\graphicspath{ {./imgs/} }

\usepackage[cmex10]{amsmath}

\usepackage{algpseudocode}
\usepackage{algorithm}
\makeatletter
\newcounter{parentalgorithm}
\newenvironment{subalgorithms}{%
  \refstepcounter{algorithm}%
  \protected@edef\theparentalgorithm{\thealgorithm}%
  \setcounter{parentalgorithm}{\value{algorithm}}%
  \setcounter{algorithm}{0}%
  \def\thealgorithm{\theparentalgorithm\alph{algorithm}}%
  \ignorespaces
}{%
  \setcounter{algorithm}{\value{parentalgorithm}}%
  \ignorespacesafterend
}
\makeatother

\usepackage[font=footnotesize]{caption}
\usepackage{array}
\usepackage{mdwmath}
\usepackage{mdwtab}
\usepackage{eqparbox}

\ifCLASSOPTIONcompsoc
\usepackage[caption=false,font=normalsize,labelfont=sf,textfont=sf]{subfig}
\else
\usepackage[caption=false,font=footnotesize]{subfig}
\fi

\usepackage{url}

\usepackage{cite}

\usepackage{multirow}
\usepackage{color}
\usepackage{breqn}
\usepackage{dsfont}
\usepackage{mathrsfs}
\usepackage[T1]{fontenc}
\usepackage{paralist}
\usepackage{booktabs}
\usepackage{threeparttable}

\newtheorem{theorem}{Theorem}[section]
\newtheorem{lemma}[theorem]{Lemma}
\newtheorem{proposition}[theorem]{Proposition}

\newenvironment{definition}[1][Definition.]{\begin{trivlist}
\item[\hskip \labelsep {\bfseries #1}]}{\end{trivlist}}

\newenvironment{remark}[1][Remark.]{\begin{trivlist}
\item[\hskip \labelsep {\bfseries #1}]}{\end{trivlist}}

\def\CRm{\mathcal C_-}
\def\CRp{\mathcal C_+}

\usepackage{enumerate}
\usepackage{setspace}

\usepackage{tikz}
\def\checkmark{\tikz\fill[scale=0.4](0,.35) -- (.25,0) -- (1,.7) -- (.25,.15) -- cycle;}

\makeatletter
\def\ps@headings{%
\def\@oddhead{\mbox{}\scriptsize\rightmark \hfil \thepage}%
\def\@evenhead{\scriptsize\thepage \hfil \leftmark\mbox{}}%
\def\@oddfoot{}%
\def\@evenfoot{}}
\makeatother
\DeclareMathOperator*{\argmax}{\arg\!\max}

\begin{document}
\title{Distributed Association and Relaying with Fairness in Millimeter Wave Networks}
\author{Yuzhe Xu,~\IEEEmembership{Student Member,~IEEE,} Hossein Shokri-Ghadikolaei,~\IEEEmembership{Student Member,~IEEE,} and Carlo Fischione,~\IEEEmembership{Member,~IEEE}
\thanks{The authors are with Electrical Engineering School and Access Linnaeus Center, KTH Royal Institute of Technology, Stockholm, Sweden. (emails: \ttfamily{\{yuzhe, hshokri, carlofi\}@kth.se}).}
\thanks{This work was supported by the Swedish Research Council.}}
\maketitle
\vspace{-1.5cm}
\begin{abstract}
Millimeter wave (mmWave) systems are emerging as an essential technology to enable extremely high data rate wireless communications. The main limiting factors of mmWave systems are blockage (high penetration loss) and deafness (misalignment between the beams of the transmitter and receiver). To alleviate these problems, it is imperative to incorporate efficient association and relaying between terminals and access points. Unfortunately, the existing association techniques are designed for the traditional interference-limited networks, and thus are highly suboptimal for mmWave communications due to narrow-beam operations and the resulting non-negligible interference-free behavior. This paper introduces a distributed approach that solves the joint association and relaying problem in mmWave networks considering the load balancing at access points. The problem is posed as a novel stochastic optimization problem, which is solved by distributed auction algorithms where the clients and relays act asynchronously to achieve optimal client-relay-access point association. It is shown that the algorithms provably converge to a solution that maximizes the aggregate logarithmic utility within a desired bound. Numerical results allow to quantify the performance enhancements introduced by the relays, and the substantial improvements of the network throughput and fairness among the clients by the proposed association method as compared to standard approaches. {It is concluded that mmWave communications with proper association and relaying mechanisms can support extremely high data rates, connection reliability, and fairness among the clients.}
\end{abstract}
\vspace{-0.5cm}

\section{Introduction}\label{sec: introduction}
Increased demands for higher data rates, along with new applications such as massive wireless access, and limited available spectrum below 6~GHz have motivated enhancing spectral efficiency by using advanced technologies such as full-duplex communications, cognitive and cooperative networking, interference cancelation, and massive multiple input multiple output (MIMO). As these enhancements are reaching their fundamental limitations, the millimeter wave (mmWave) band is becoming an alternative and promising option to support extremely high data rate wireless access~\cite{Rappaport2013Millimeter,Rangan2014Millimeter,shokri2015mmWavecellular,Niu2015Survey,rappaport2014mmWaveBook}. There is a growing consensus in both academia and industry that mmWave communication technology will play an important role in next generation wireless networks. Motivated by future demands for high data rates, several standardization activities within wireless personal area networks (WPANs) and wireless local area networks (WLANs) have been undertaken; examples include IEEE~802.15.3c~\cite{802_15_3c}, IEEE~802.11ad~\cite{802_11ad}, ECMA~387~\cite{ECMA387}, WirelessHD consortium, the wireless gigabit alliance, and recently IEEE~802.11ay.\footnote{Detailed information can be found at \url{http://www.wirelesshd.org}, \url{http://wirelessgigabitalliance.org}, and \url{http://www.ieee802.org/11/Reports/ng60_update.htm}, respectively.}

The main challenges of mmWave networks are severe path-loss (distance-dependent component of channel attenuation), deafness, and blockage~\cite{Rappaport2013Millimeter,Rangan2014Millimeter,shokri2015mmWavecellular,Niu2015Survey,rappaport2014mmWaveBook}. To compensate for the high path-loss, the small wavelengths of mmWave frequencies allow for the implementation of a large number of antenna elements in the current size of radio chips to make narrow beams (known as pencil-beams). These narrow beams provide significant antenna gains, which boost the link budget. Narrow beams also substantially reduce the interference footprint, as the receiver only listens to a specific direction, which reduces the effective number of transmitters generating the interference~\cite{shokri2015mmWavecellular,Singh2011Interference}. In fact, narrow beams result in frequent situations in which multiuser interference is negligible~\cite{Shokri2016Transitional}, and we may face a noise-limited network as opposed to the conventional interference-limited microwave networks. {However, connection establishment and maintenance with narrow beams imposes a time consuming alignment overhead~\cite{Shokri2015Beam} to avoid \emph{deafness}, i.e., a situation in which the main beams of the transmitter and the receiver do not point to each other, making it impossible to establish a high quality mmWave link. }An additional challenge of mmWave {communications} is \emph{blockage}, i.e., high penetration loss. To have quantitative insights, the human body can attenuate mmWave signals by 35~dB~\cite{lu2012modeling}, and materials as brick attenuate by as much as 80~dB~\cite{allen1994building,alejos2008measurement,Rangan2014Millimeter}.

The problems mentioned above cannot be efficiently solved by just increasing the transmit power or by adding antenna gain using narrower beams~\cite{rappaport2014mmWaveBook}, but only by re-association or relaying procedures.\footnote{Alternatively, a blockage can be addressed by using reflections. However, the existence of such reflectors with sufficiently large reflection indices is not guaranteed in all environments. Moreover, there will be always some additional loss due to any reflection, which may be intolerable for high quality mmWave links.} The association and relaying is particularly important in mmWave networks due to the limited size of the cells and dense access points (AP) deployment~\cite{Singh2009Blockage,kim2013joint}. Relaying techniques can provide more uniform quality of service by offering robust mmWave connection, load balancing, coverage extension, indoor-outdoor coverage, efficient mobility management, and smooth handover operation~\cite{Rangan2014Millimeter,rappaport2014mmWaveBook,Niu2015Survey,shokri2015mmWavecellular,Singh2009Blockage,kim2013joint}. In~\cite{Singh2009Blockage}, it is shown that having an alternative path using relays in mmWave networks can increase the connectivity by about 100\%. Further, extensive analysis in~\cite{kim2013joint} demonstrates that relays can effectively extend the range to support high quality live video streaming over $300$\,m. Therefore, in mmWave networks, the association of a client or user to an AP (or base station) and relaying are some of the first and most important routines.

Given the association may govern the long-term resource allocation policies of conventional wireless networks~\cite{andrews2014overview}, it has been the focus of intense research in the last years\cite{andrews2014overview,Bejerano2007,Shakkottai2006,Kauffmann2007,ye2013user,Lee2007,Athanasiou2009,Athanasiou2014Auction,Athanasiou2015,Shen2014,Lin2013,Boostanimehr2015}. The current mmWave standards use the minimum-distance association, which leads to a simple association metric based on the received signal strength indicator (RSSI)~\cite{802_15_3c,802_11ad,ECMA387}. Although RSSI association metrics are suitable for an interference-limited homogenous network, they may lead to poor use of the available resources in the presence of non interference-limited environments, non-uniform spatial distribution of clients, and heterogenous APs/relays with a different number of antenna elements and different transmission powers~\cite{andrews2014overview}. These standardized association approaches lead to an unbalanced number of clients per AP, which limits the available resources per client in highly populated areas~\cite{shokri2015mmWavecellular}, while wasting resources in sparse areas. This poor load balancing indeed decreases network-wide fairness, since overloaded APs cannot provide their associated clients as much resource as less-loaded APs. Thus, it is possible for the clients to associate with father APs for the better load sharing, without suffering from a huge path loss drop.

Beside mmWave association techniques from the standards, there are many more solutions for association and relaying from the literature of microwaves networks. In~\cite{Bejerano2007}, a client association policy is investigated to ensure network-wide max-min fair bandwidth allocation to the clients in WLANs. The network throughput maximization using load balancing is proposed in~\cite{Shakkottai2006} by using a fluid model of client population. \cite{Kauffmann2007} presents a set of self-configuring algorithms using Gibbs sampler to improve association and fair resource sharing without explicit coordination among the wireless devices. In~\cite{Lee2007}, a dual-association approach in wireless mesh networks is presented, where the APs for unicast traffic and those for broadcast traffic are independently chosen by exploiting overlapping coverage and maximizing the unicast throughput. In the seminal work of~\cite{ye2013user}, a joint association and resource allocation problem is formulated for a heterogenous cellular network. The authors ensure network-wide proportional fairness via a logarithmic utility maximization and propose a distributed solution approach via dual decomposition. Dynamic association and reassociation procedures are introduced in~\cite{Athanasiou2009}. The procedures use the uplink/downlink channel conditions and the traffic load in the network.

Unfortunately, the association procedures of the literature above are highly sub-optimal for mmWave networks due to frequent handovers caused by\begin{inparaenum}[\itshape i\upshape)]
\item dense deployment of APs,
\item vulnerability to random obstacles,
\item deafness, and
\item loss of precise beamforming information due to channel changes~\cite{shokri2015mmWavecellular}.
\end{inparaenum}Moreover, the major limitation of the aforementioned approaches is that they are primarily designed for interference-limited microwave networks, where the interference level hinders the benefits of load balancing for networks with ultra dense APs (base stations) deployment~\cite{andrews2014overview}. However, interference is not the major limitation of mmWave networks. The high directionality level in mmWave both at the transmitter and at the receiver is a distinguishing feature, which may provide a new interference footprint: a noise-limited regime~\cite{shokri2015mmWavecellular}. These fundamental differences between mmWave networks and the conventional microwave ones demand novel association metrics and procedures. Reducing the overhead of frequent reassociation, together with the natural need of load balancing among the APs, justifies that a client in mmWave networks may be advantageously served by a farther but less-loaded and easy-to-find AP~\cite{shokri2015mmWavecellular}. Robustness of the association to random blockage should be improved to reduce the number, and thereby the overhead/delay, of reassociation and to provide a seamless handover~\cite{rappaport2014mmWaveBook,shokri2015mmWavecellular}.

We can roughly adopt two association options in mmWave networks to improve the robustness to blockage:
\begin{inparaenum}[\itshape i\upshape)]
\item multiple parallel connectivity, and
\item single sequential connectivity~\cite{shokri2015mmWavecellular}.
\end{inparaenum}
In the first approach, a client adopts multi-beam transmissions toward several APs (relays) at the same time to establish multiple paths. This approach provides seamless handover, robustness to blockage, and continuous connectivity. The prices are an SNR loss for each beam if we consider a fixed total power budget for every transmitter\footnote{Advanced joint scheduling and beamforming strategies such as CoMP may compensate for this SNR loss. The required signaling overhead, however, may be overwhelming in mmWave networks~\cite{shokri2015mmWavecellular}.}, more complicated resource management and relay selection, and higher signaling and computational complexities for beamforming. To enable multi-beam operation, we may need to have hybrid beamforming~\cite{sun2014}.\footnote{Complete digital beamforming with high resolution analog-to-digital-converters may be impractical in mmWave networks with huge bandwidth and with massive number of antenna elements both at the transmitter and at the receiver.} To alleviate computational and signaling overhead of the beamforming with many antenna elements, current mmWave standards adopt an analog beamforming~\cite{rappaport2014mmWaveBook}. This simplification comes at the expense of having only antenna gain without any multiplexing gain, as every client can make only one beam, avoiding realization of multiple parallel connectivity. Instead, a client may be associated with several APs (relays) with several paths, but the connection will be established using only one of these paths at a time, whereas the others are used as backup. This single sequential connectivity scenario is standard-compliant and mitigates disadvantages of the multiple parallel connectivity scenario, see~\cite{shokri2015mmWavecellular} for detailed comparisons. In this paper, we focus on the single sequential connectivity.

Our previous approaches~\cite{Athanasiou2014Auction,Athanasiou2015} were among the first studies to address the association problem in 60~GHz mmWave communications. However, those approaches did not consider relays, a vital part of mmWave networks, which substantially increases the difficulty of the association and relaying problem. Relay selection, per se, has a rich literature~\cite{Abdulhadi2012Relay} in the context of cooperative communications. To name relevant works, an amplify-and-forward wireless relay networks is considered in~\cite{Cai2008}, in which a semi-distributed algorithm on energy-efficient relay node selection is proposed for a multiple-source multiple-destination scenario, but with no strong optimality guarantees. In~\cite{ubaidulla2012}, a two-way relaying scheme is established for cognitive radio networks, where only a pair of secondary transceiver nodes is considered to communicate with each other assisted by a set of cognitive two-way relays. However, directional communications with narrow beams make it very hard to apply these relaying techniques such as amplify-and-forward to decode-and-forward, where the transmitted signal from direct path and from the relay path is superimposed at the receiver. Instead, as commonly assumed in the mmWave literature~\cite{Singh2007,Singh2009Blockage,Huang2008}, a receiver can receive signal either from the direct path or from the relay path, not both. The relays therefore act as virtual APs. In particular, once a client transmits the signal to a relay, then the relay forwards the signal together with its own signal to an AP. \cite{Singh2007,Singh2009Blockage} propose a cross-layer approach to select either direct transmission mode (without relays) or two-hop mode (with a random relay selection). The decision criteria is only the existence of line-of-sight (LoS) on the direct link. In~\cite{Huang2008}, an auction-based resource allocation algorithm is proposed for relay selection to provide max-min fairness among the clients.

\begin{table*}[t]
  \centering
  \begin{threeparttable}[b]
  \caption{Summary of the existing solutions for association and relay selection problems.\tnote{1}}\label{tab:comparison literature}
  \scriptsize
  \renewcommand{\arraystretch}{0.9}
\begin{tabular}{lccccccccccc}
  \toprule
   & Assoc. & Alloc. & Relay  & Stoch. & Through. & Fair. & Backup & {Network} & Distribut.  & Optim.  & Complex. \\
  \midrule
  \cite{Bejerano2007} & \checkmark & - & - & - & \checkmark & \checkmark & - & {WLAN} & - & \checkmark & -\\\midrule
  \cite{Kauffmann2007}& \checkmark & \checkmark & - & - & \checkmark & \checkmark & - & {WLAN} & \checkmark & local & - \\\midrule
  \cite{ye2013user} & \checkmark & \checkmark & - & - & \checkmark & \checkmark & - & {HetNet} & \checkmark & - & - \\\midrule
  \cite{Athanasiou2009} & \checkmark & - & - & -& \checkmark & \checkmark & - & {WLAN} & \checkmark & - & - \\\midrule
  \cite{Athanasiou2014Auction} & \checkmark & - & \checkmark & - & - & - & - & {mmWave} & \checkmark & \checkmark & -\\\midrule
  \cite{Athanasiou2015} & \checkmark & - & \checkmark & - & - & \checkmark & - & {mmWave} & \checkmark & \checkmark & - \\\midrule
  \cite{Shen2014} & \checkmark & \checkmark & - & - & \checkmark & \checkmark & - & {HetNet} & \checkmark & {\checkmark} & \checkmark\\\midrule
  \cite{Lin2013} & \checkmark & \checkmark & - & \checkmark & \checkmark & \checkmark & -& {HetNet} & offline & \checkmark & - \\\midrule
  \cite{Boostanimehr2015} & \checkmark & \checkmark & - & - & \checkmark & \checkmark & - & {HetNet} & \checkmark & - & - \\\midrule
  \cite{Cai2008} & - & \checkmark & \checkmark & - & \checkmark & - & - & {WRN} & semi & - & -  \\\midrule
  \cite{Huang2008} & - & \checkmark & \checkmark & - & \checkmark & \checkmark & - & {WRN} & \checkmark & - & - \\
  \midrule
  {\color{blue}\cite{Bu2006} }& {\color{blue}\checkmark} & {\color{blue}-} & {\color{blue}-} & {\color{blue}-} & {\color{blue}\checkmark}
   & {\color{blue}\checkmark }&   {\color{blue}-} &{\color{blue} {WLAN}} & {\color{blue}-} & {\color{blue}\checkmark }& {\color{blue}-} \\\midrule
  {\color{blue}\cite{Prasad2014}}& {\color{blue}\checkmark} &{\color{blue} \checkmark} & {\color{blue}-} &{\color{blue} -} & {\color{blue}\checkmark }& {\color{blue}\checkmark} & {\color{blue}- }& {\color{blue}{HetNet}} & {\color{blue}-} & {\color{blue}{\checkmark}} &{\color{blue} \checkmark}\\\midrule
  Ours & \checkmark & \checkmark & \checkmark & \checkmark & \checkmark & \checkmark & \checkmark & {mmWave} & \checkmark & \checkmark &\checkmark\\
  \bottomrule
\end{tabular}
\begin{tablenotes}
\scriptsize
\item [1] ``Assoc'', ``Alloc'', and ``Relay'' represent association, resource allocation (including power allocation), and relay selection, respectively. Stoch. indicates stochastic optimization problems. ``Through'' and ``Fair'' are network throughput and fairness. ``Backup'' indicates backup associations. {``Network'' characterizes different networks: wireless local area network (``WLAN''), heterogeneous network (``HetNet''), wireless relay network or cooperative network (``WRN''), and ``mmWave'' networks.} ``Distribut'' and ``Optim'' indicate that the solution approaches are distributed and optimal (or near optimal) guaranteeing, respectively. ``Complex'' represents computational complexity analysis. A check mark in each column indicates that the feature is considered.
\end{tablenotes}
\end{threeparttable}
\vspace{-0.5cm}
\end{table*}

None of the previous approaches studied jointly the association and relaying problems in mmWave networks with possible negligible multiuser interference. Random relay selection, as proposed in~\cite{Singh2009Blockage,Singh2007}, while improves the robustness to blockage, may lead to significant throughput drop in a cooperative network~\cite{ubaidulla2012}. In fact, relay selection affects heavily the ability of a terminal to reach a farther AP and the interference footprint of the network, and also determines how the available resources should be distributed among the clients (resource allocation). As an association problem can be transformed into a long-term resource allocation instance~\cite{andrews2014overview}, association and relaying are strongly interconnected and a joint solution, where clients are also candidates for relaying other client's traffic is of great importance in mmWave wireless networks. {Table~\ref{tab:comparison literature} summarizes existing solutions for association and relay selection problems. The purpose of the table is to highlight the different characteristics of this paper and the existing literature.}

In this paper, we investigate the joint association and relaying problem for mmWave networks.\footnote{We mainly focus on mmWave WPANs and WLANs throughout the paper, and the results obtained can be readily extended to mmWave cellular networks.} Our main contributions are summarized as follows:
\begin{enumerate}[\itshape i\upshape)]
\item We propose a novel optimization approach to the joint association and relaying problem, in which a logarithmic utility, resource allocation for APs, {association for clients, relay selection, and imperfect channel state information are considered all together. Given a fixed client association and relaying, we find the closed-form optimal rates and resource allocations for clients and APs, respectively.}
\item {We further reformulate the joint association and relaying problem above as a multi-dimensional assignment problem, for which we propose a novel solution approach. }The approach is inspired by dual-primal decomposition, which under special conditions allows the {multi-dimensional} assignment problem to be safely relaxed in the binary constraint, and still to have the same optimal objective value.
\item We establish a distributed association algorithm based on a novel distributed auction algorithm to solve the {multi-dimensional} assignment problem. We systematically investigate the convergence properties of the proposed distributed auction algorithm, and show that it provides a near optimal solution in polynomial time.
\end{enumerate}

The rest of the paper is organized as follows: In Section~\ref{sec:system-model}, we introduce the system model and formulate the joint client association and relaying problems. In Sections~\ref{sec:centralized-approach} and \ref{sec:distributed-approach}, we propose our centralized and distributed solution approaches, respectively. Numerical results are reported in Section~\ref{sec:numerical-examples}, and our conclusions follow in Section~\ref{sec:conclusion}.

\vspace{-0.5cm}
\section{System Model And Problem Formulation}
\label{sec:system-model}

\begin{table}[t]
  \centering
  \caption{\color{blue}Main Notations}
  \label{tab_symbol}
  \scriptsize
  {\color{blue}
  \begin{tabular}{ll}
    \toprule
    Symbols & Description \\
    \midrule
    $\mathcal A$    & set of APs \\
    $\mathcal C$  & set of clients \\
    $\mathcal C_+ \subseteq \mathcal C$         & set of clients with relaying capability \\
    $\mathcal C_- = \mathcal C \setminus \mathcal C_+$ &  set of clients without relaying capability\\
    $c_{ij}$ & achievable rate between entities $i$ and $j$\\
    $\mathbf r$ & (peak) rates\\
    $\mathbf x$ & binary association indicator\\
    $\mathbf y$ & resource fraction\\
    $n_k$ & number of clients connecting to AP $k$\\
    $\epsilon$ & desired accuracy for auction algorithms\\
    \bottomrule
  \end{tabular}}
\end{table}

\begin{figure}[t]
  \centering
  \includegraphics[width=0.4\textwidth]{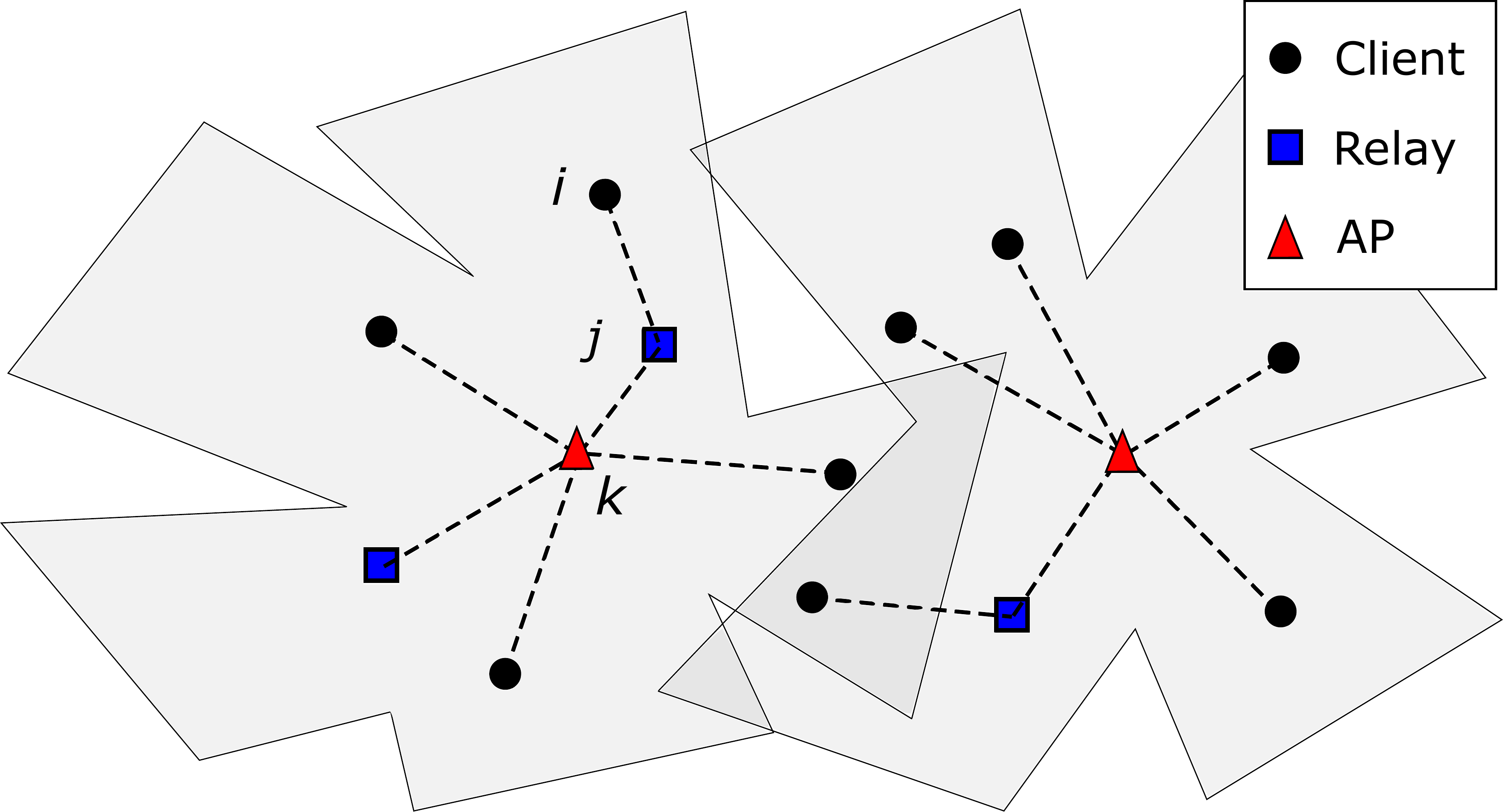}
  \caption{An example network with 9 clients, 3 relays, and 2 APs. Irregular shapes in grey show the serving regions of the APs. We do not have regular circular/hexagonal serving regions in mmWave networks due to blockage, pencil-beam operation, and load balancing~\cite{shokri2015mmWavecellular}. } \label{fig network}
  \vspace{-0.5cm}
\end{figure}

Denote by $\mathcal A$ and $\mathcal C$ the set of APs and clients, respectively. The clients must be associated with one of the available APs to establish a communication. Moreover, each client may skip the direct association to an AP and establish a connection via one of the available relays, where a relay is a client that can help other clients to be served by APs in addition to its own transmissions~\cite{Ding2011}. {Denote by $\CRp \subseteq \mathcal C$ the set of the relays (clients with relaying capability), and define set $\CRm = \mathcal C \setminus \CRp$ for the rest clients (clients without relaying capability). In the rest of paper, to avoid heavy notation, we use ``client'' and ``relay'' to name the entities in set $\CRm$ and $\CRp$, respectively, when not stated in the description.} Moreover, denote by $c_{ij}$ the achievable rate between mmWave entities $i$ and $j$. We let $c_{ij}=0$, if there is an obstacle between entities $i$ and $j$~\cite{rappaport2014mmWaveBook}. An example of the network is illustrated in Fig.~\ref{fig network}. Note that in mmWave networks, the vulnerability due to obstacles, high level of directionality, and load balancing, along with noise-limited operation, results in irregular serving regions of APs~\cite{shokri2015mmWavecellular,Andrews2014What}, as highlighted in Fig.~\ref{fig network}. In fact, being served by the closest AP, exemplified by circular/hexagonal serving regions, is not a proper option (may be impossible due to blockage) in mmWave networks.

Each entity of the network (such as client, relay, and AP) is equipped with steerable directional antennas to make the typical pencil-beams of mmWave communications. In this paper, we adopt the single sequential connectivity scenario and assume that every client can make only one beam, as standardized in~\cite{802_11ad,802_15_3c}, due to the huge complexity of making several beams. Moreover, an entity is only able to transmit to (receive from) another entity. This is {compatible} with the existing mmWave standards~\cite{802_11ad,802_15_3c}, where unicast is mainly supported and virtual broadcast (e.g., synchronization) is realized by scanning all directions. Extension to multiple parallel connectivity using multi-beam operation, which is a complex and promising approach for mmWave cellular networks, is left for future studies.

We further assume that clients operate in the half-duplex mode, whereas relays can operate in the full-duplex mode, which is feasible for short-range mmWave networks as stated in~\cite{Liangbin2014,Rajagopal2014} and promising for cellular networks. The required modifications for half-duplex relays is straightforward and provided in~\cite{kim2013joint}. Due to the validity of the negligible multiuser interference assumption in small to modest-size networks~\cite{Singh2011Interference,Shokri2016Transitional}, we may ignore the interference and approximate the achievable rates of individual link between client $i\in\CRm$ and AP $k\in\mathcal A$ as
\begin{align}\label{eq:capacity-model}
    c_{ik} = \log_2 \left( 1 + {\rm SNR}_{ik} \right)\,,
\end{align}
where ${\rm SNR}_{ik}$ is the signal-to-noise ratio of the link between client $i$ and AP $k$. Similarly, we define achievable rates $c_{ij}$ and $c_{jk}$ for $i\in\CRm$, $j\in\CRp$, and $k\in\mathcal A$. We also assume that the alignment overhead is negligible while formulating the association and resource allocation problem; however, we use backup connections, as we will see later in next sections, to alleviate frequent handover and re-association problem. In~\cite{Shokri2015Beam}, a performance evaluation framework is developed to identify the alignment-throughput tradeoff and to find the required directionality level in mmWave networks. Using a similar framework, we could introduce the alignment overhead into the formulations of this paper, which will be undertaken in our future studies.

\vspace{-0.5cm}
\subsection{Logarithmic Utility Maximization}
\label{subsec:gener-utili}

Let real variables $r_i$ and $r_j$ be the rates of client $i$ and relay $j$, respectively. Denote by $y_{ik}$ the fraction of the resources allocated to serve client $i$ in AP $k$. Let $x_{ik}$, $x_{ij}$, and $x_{jk}$ be binary association indicators. In particular, $x_{ik} = 1$ if client $i$ is associated with AP $k$, otherwise $x_{ik} = 0$. $x_{ij} x_{jk} =1$ if client $i$ is served by AP $k$ via relay $j$. Thus, $r_i x_{ik}$ is the rate of client $i$ toward AP $k$ once AP $k$ grants the direct channel access to this client (peak rate), and $r_i y_{ik}$ is the effective (average) rate of client $i$ if it is served by AP $k$.

Note that the peak rate of every link must be lower than the link capacity. Moreover, each relay $j$ transmits its own data at rate $r_j$ and may opportunistically assist one client. Therefore, from each client $i$ to each relay $j$ or to each AP $k$, we have
\begin{align}\label{eq:const-linkcap-i}
r_i x_{ij} &\leq c_{ij} & r_i x_{ik} &\leq c_{ik} & \forall i\in\CRm,j\in\CRp, k\in\mathcal A\,,
\end{align}
and from each relay $j$ to each AP $k$, {considering superposition coding, }we have
\begin{align}
\label{eq:const-linkcap-j}
\left(\sum_{i\in\CRm} r_{i}x_{ij} + r_{j}\right)x_{jk} &\leq c_{jk}&& \forall j \in \CRp, k\in\mathcal A\,.
\end{align}
Constraints~\eqref{eq:const-linkcap-i} and \eqref{eq:const-linkcap-j} require precise SNR values to determine the achievable rates (channel capacities) by \eqref{eq:capacity-model}. However, in practical systems, a client/AP cannot have precise SNR values due to imperfect channel estimation and also limited feedback channel~\cite{love2008}, which affects the general design principles of mmWave systems~\cite{el2014spatially}, as well as the relay selection performance~\cite{Ding2011}. In particular, we assume that a client estimates SNR values, namely $\widetilde{\textrm{SNR}}_{ij}$ for the individual link between client $i$ and relay $j$, which is defined as
\begin{align*}
  \widetilde{\textrm{SNR}}_{ij} = \textrm{SNR}_{ij} + e_{ij}\,,
\end{align*}
where $e_{ij}$ is the error due to estimation and limited feedback channel. The statistical properties of the error $e_{ij}$ can be derived using the similar technique as in~\cite{el2014spatially}. In this paper, we assume that the statistical knowledge of $e_{ij}$ is available a priori. Therefore, we have the cumulative distribution function of SNR given its corresponding estimates at clients, namely we have
\begin{align}
\label{eq:cdf-estimate}
  F^{\gamma}_{ij}(z) & = \Pr \left(\textrm{SNR}_{ij}\leq z\, | \widetilde{\textrm{SNR}}_{ij}= \gamma\right) = \int_{\gamma - z}^\infty f_{e_{ij}}(x) dx\,,
\end{align}
where $f_{e_{ij}}$ is the probability density function of the estimation error $e_{ij}$. For example, suppose we have $F_{ij}^{10}(10) = 0.9$, which implies that if client $i$ obtains $\widetilde{\textrm{SNR}}_{ij}=10\,$dB for the link toward relay $j$, then we have the probability of SNR$_{ij}\leq 10\,$dB is 90\% {from \eqref{eq:cdf-estimate}}. Considering both \eqref{eq:capacity-model} and \eqref{eq:cdf-estimate}, for $i\in\CRm$, $j\in\CRp$ and $k\in\mathcal A$, constraint~\eqref{eq:const-linkcap-i} becomes
\begin{align}
\label{eq:const-stoch-linkcap-ij}
\Pr \left( r_i x_{ij} \geq c_{ij} | \widetilde{\textrm{SNR}}_{ij} \right) &\leq \eta, \\
\label{eq:const-stoch-linkcap-jk}
\Pr \left( r_i x_{ik} \geq c_{ik} | \widetilde{\textrm{SNR}}_{ik} \right) &\leq \eta\,,
\end{align}
where $\eta$ is the design parameter. Similarly, for all $j\in\CRp$ and $k\in\mathcal A$, constraint~\eqref{eq:const-linkcap-j} becomes
\begin{align}
\label{eq:const-stoch-linkcap-j}
\Pr \left( \left(\sum_{i\in\CRm} r_{i}x_{ij} + r_{j}\right)x_{jk} \geq c_{jk} \Bigg|  \widetilde{\textrm{SNR}}_{jk} \right) \leq \eta\,.
\end{align}
With one beam per client, each client $i$ must be associated either with an AP or is connected to a relay, i.e.,
\begin{align}\label{eq:const-assoc-i}
\sum_{j\in \CRp} x_{ij} + \sum_{k\in \mathcal A} x_{ik} &= 1, && \forall i \in \CRm\,,
\end{align}
whereas each relay $j$ can assist at most one client, i.e.,
\begin{align}
\label{eq:const-assoc-j}
\sum_{i\in \CRm} x_{ij} &\leq 1 & \sum_{k\in\mathcal A} x_{jk} &= 1, && \forall j \in \CRp\,.
\end{align}
Furthermore, {once a relay assists a client, it sends both the client's and its own traffic over the same time-frequency-spatial resource block (see \cite{Liangbin2014,Rajagopal2014}) by superposition coding schemes, as illustrated in Fig.~\ref{fig:resource-sharing}.
\begin{figure}[t]
  \centering
  \includegraphics[width=0.33\textwidth]{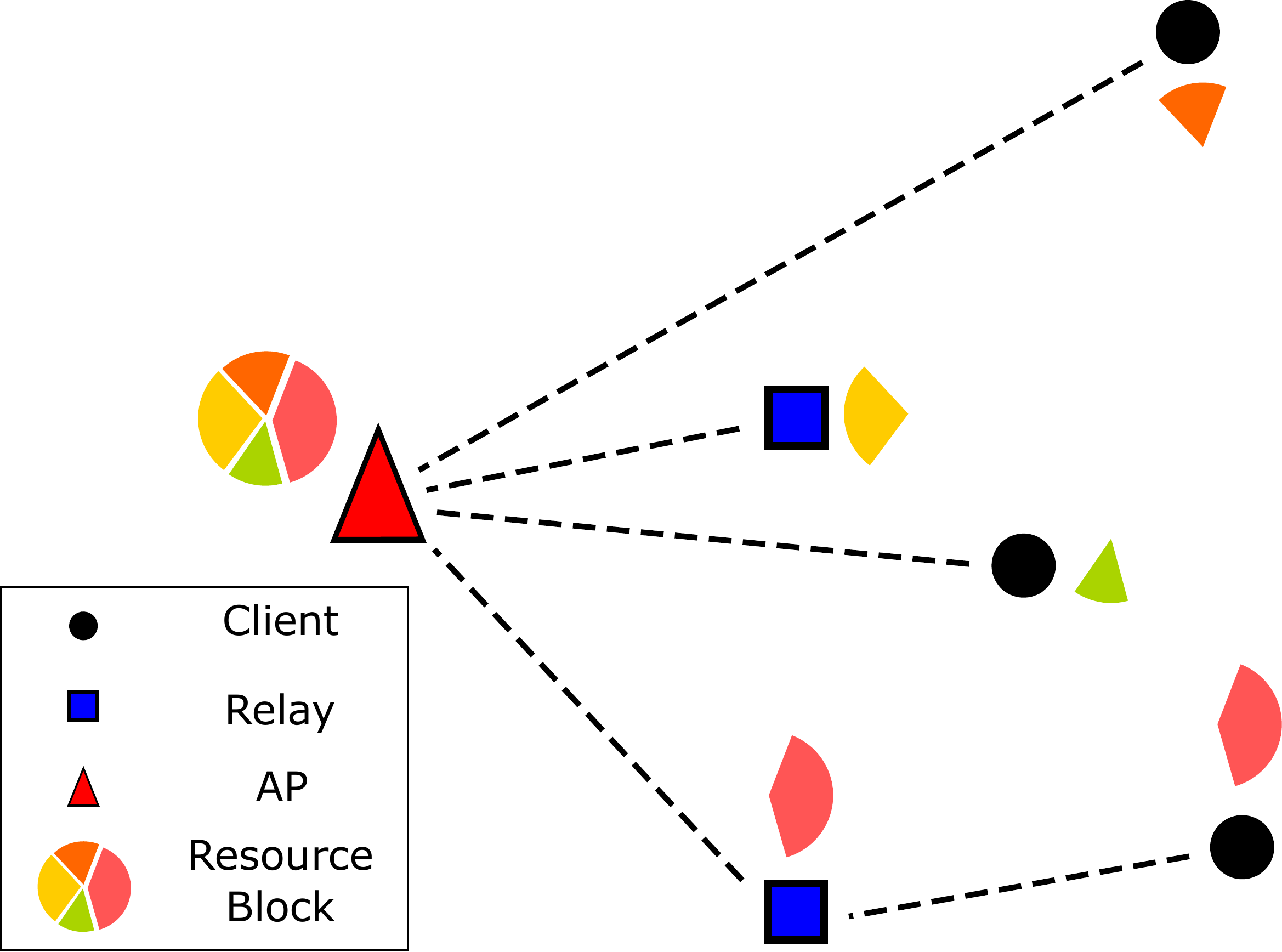}
  \caption{An illustration of the resource sharing at one AP. One sector represents a resource fraction assigned by the AP. Note that once a client is associated with a relay, then the resource fraction for this client is same as that for the relay. }\label{fig:resource-sharing}
  \vspace{-0.8cm}
\end{figure}
Therefore, the resource fraction for client $i$ is the same as that for relay $j$, if $x_{ij}=1$, i.e., }
\begin{align}\label{eq:const-fract-ij}
(y_{ik}-y_{jk})x_{ij} &= 0, && \forall i \in \CRm, j\in \CRp, k\in\mathcal A\,.
\end{align}
Last but not least, recall respectively the binary variables $x_{ij}$, $x_{ik}$ and $x_{jk}$:
\begin{align}\label{eq:const-x}
x_{ij}, x_{ik}, x_{jk} &\in \{0,1\}, && \forall i\in \CRm, j\in\CRp, k\in\mathcal A\,,
\end{align}
and the resource allocation fraction variables $y_{ik}$:
\begin{align}\label{eq:const-y}
{\sum_{i\in\mathcal C} x_{ik}y_{ik}}&\leq 1, & y_{ik}&\geq 0, && \forall i\in\mathcal C, k\in\mathcal A\,.
\end{align}
Note that, roughly speaking, association is a long-term resource allocation problem. How to allocate the long-term resource share of individual clients to guarantee certain level of delay and to handle bursty traffics are focus of short-term resource allocations, which is out of the scope of this work.

Suppose the objective is to maximize the aggregate utility function:
{
\begin{align}
\label{eq:objec}
\sum_{i\in\CRm}  f_i \left( r_i \sum_{k\in\mathcal A} y_{ik} \left( x_{ik} +  \sum_{j\in\CRp} x_{ij} x_{jk}\right) \right) +  \sum_{j\in\mathcal C_+}  f_j \left( r_j \sum_{k\in\mathcal A} y_{jk} x_{jk} \right)
\end{align}
where} utility $f_i: \mathds R \rightarrow \mathds R$ for $i\in\mathcal C$ is a continuously differentiable, monotonically increasing, and strictly concave function. These conditions hold for most of the practical utility functions, for instance, for the logarithmic utility. {The utility of client $i\in\mathcal C_-$ associated with AP $k_i$ is
\begin{align}
\label{eq:utility-client}
   f_i \left( r_i \sum_{k\in\mathcal A} y_{ik} \left( x_{ik} +  \sum_{j\in\CRp} x_{ij} x_{jk}\right) \right) = f_i (r_i y_{ik_i})\,,
\end{align}
no matter whether it is directly associated with AP $k_i$ or is connected to the AP via a relay. Moreover, defining $0:= 0 \cdot f_i(0)$ for $i\in\mathcal C$, \eqref{eq:utility-client} becomes
\begin{align*}
  &f_i (r_i y_{ik_i}) = \sum_{k\in\mathcal A} x_{ik} f_i (r_i y_{ik}) + \sum_{j\in\mathcal C_+} \sum_{k\in\mathcal A} x_{ij}x_{jk} f_i (r_i y_{ik}) \,,
  &&\forall i \in \mathcal C_-\,.
\end{align*}
With similar derivation for relay $j\in\mathcal C_+$, we have
\begin{align}
&f_j \left( r_j \sum_{k\in\mathcal A} y_{jk} x_{jk} \right) =\sum_{k\in\mathcal A} x_{jk} f_j (r_j y_{jk})\,, && \forall j\in\mathcal C_+\,.
\end{align}
Then, \eqref{eq:objec} becomes}
\begin{align}
\label{eq:objec-1}
  {
  \sum_{i\in\mathcal C} \sum_{k\in\mathcal A} x_{ik} f_i (r_i y_{ik}) + \sum_{i\in\mathcal C_-} \sum_{j\in\mathcal C_+} \sum_{k\in\mathcal A} x_{ij}x_{jk} f_i (r_i y_{ik})\,,
  }
\end{align}
{due to binary variables $x_{ik}$, $x_{ij}$, $x_{jk}$, and constraints~\eqref{eq:const-assoc-i}, \eqref{eq:const-assoc-j}. The first term in \eqref{eq:objec-1} is the aggregate utility of both the clients associated directly with APs, and the relays, whereas the second term is that of the clients assisted by relays. Using a logarithmic utility function, which is naturally used for load balancing and proportional fairness provisioning among the clients~\cite{andrews2014overview}, the resulting objective function \eqref{eq:objec-1} becomes}
\begin{align}
\label{eq.objective-log}
{
	\sum_{i\in\mathcal C}\sum_{k\in\mathcal A} x_{ik}\log \left( r_i y_{ik} \right) + \sum_{i\in\mathcal C_-} \sum_{j\in\mathcal C_+} \sum_{k\in\mathcal A} x_{ij}x_{jk} \log (r_i y_{ik}) \,,
}
\end{align}
where {we define $0:=0\cdot \log 0 $, which is the limit of $u\log u$ as $u$ approaches to 0}. Note that the logarithm utility encourages to the proportional fairness among the clients, which naturally satisfies a realistic resource allocation strategy.

Let $\mathbf x$, $\mathbf r$, and $\mathbf y$ denote the vectors that collect association indicators $x_{ij}$, $x_{ik}$ and $x_{jk}$, the rates $r_i$, and the fraction variables $y_{ik}$, respectively. Considering constraints~\eqref{eq:const-stoch-linkcap-ij}$\sim$\eqref{eq:const-y}, we can pose the following stochastic optimization problem to jointly optimize the association and resource allocation in mmWave networks:
\begin{align}\label{eq:stoch-optim-maxflow}
    \underset{\mathbf x,\:\mathbf r,\:\mathbf y}{\textrm{max}} & \quad {
	\sum_{i\in\mathcal C}\sum_{k\in\mathcal A} x_{ik}\log \left( r_i y_{ik} \right) + \sum_{i\in\mathcal C_-} \sum_{j\in\mathcal C_+} \sum_{k\in\mathcal A} x_{ij}x_{jk} \log (r_i y_{ik})} &\rm{s.t.} & \quad \eqref{eq:const-stoch-linkcap-ij}\sim\eqref{eq:const-y}\,,
\end{align}
where stochastic constraints~\eqref{eq:const-stoch-linkcap-ij}$\sim$\eqref{eq:const-stoch-linkcap-j} are considered. Problem~\eqref{eq:stoch-optim-maxflow} is a stochastic and mixed-integer optimization problem. {Moreover, as shown in Section~\ref{subsec:multi-assign}, \eqref{eq:stoch-optim-maxflow} can be transformed into a multi-dimensional assignment problem with nonlinear objective function, thus it
is in general NP-hard~\cite{Bertsekas1998}}. In this paper, our main contributions {are proposing} solution approaches for optimization problem~\eqref{eq:stoch-optim-maxflow}.

{\color{blue}
In this paper, we suppose that a non-relaying-capable client $i\in\mathcal C_-$ can make use of one relay, which simplifies the analysis and the description of the solution algorithm. However, formulation~\eqref{eq:stoch-optim-maxflow} can be easily modified to include the case where a relay makes use of another relay, which is given by the following problem formulation
\begin{align}
\label{eq:relay-relay}
\underset{\mathbf x,\:\mathbf r,\:\mathbf y}{\textrm{max}} & \quad {
	\sum_{i\in\mathcal C}\sum_{k\in\mathcal A} x_{ik}\log \left( r_i y_{ik} \right) + \sum_{i\in\mathcal C} \sum_{j\neq i, j\in\mathcal C_+} \sum_{k\in\mathcal A} x_{ij}x_{jk} \log (r_i y_{ik})} &\rm{s.t.} & \quad \eqref{eq:const-stoch-linkcap-ij}\sim\eqref{eq:const-y}\,.
\end{align}
To solve~\eqref{eq:relay-relay}, it is straightforward to use the solution algorithm for~\eqref{eq:stoch-optim-maxflow}, which we present later as Algorithm~\ref{alg.primal-dual}. 
}
\vspace{-0.5cm}
\subsection{Multiple Associations}

To avoid frequent handovers and reduce the overhead/delay of reassociation, we let each client reserve a backup association with APs. {This backup path is activated only upon vanishing the original path, and therefore it requires periodical checking of the availability of the backup paths. The overhead of this signaling is much less than the overhead of periodical pilot transmissions for channel estimation and for beam-training. }To establish the backup path, we propose a two-step procedure.
{
\begin{enumerate}
  \item In the first step, we solve optimization problem~\eqref{eq:stoch-optim-maxflow} by the methods proposed later in the paper. Denote by $\boldsymbol \tau^*$ its optimal association.
  \item In the second step, we solve another optimization problem, which uses $\boldsymbol\tau^*$ as input, and aims to maximize the average number of clients having LoS at the backup connection. Thus, its objective function is given as
\begin{align}
  \label{eq:bk-object}
  \sum_{i\in\mathcal C} \sum_{k\in\mathcal A} \omega_{ik} x_{ik} + \sum_{i\in\CRm}\sum_{j\in\CRp} \sum_{k\in\mathcal A} \omega_{ij} \omega_{jk} x_{ij} x_{jk},
\end{align}
where $\omega_{ij} = q_{ij} (1-\tau_{ij}^*)$, in which $q_{ij}$ is the probability of having LoS from client $i$ to relay $j$. These probabilities can be characterized by experimental~\cite{Schaubach1992} and analytical~\cite{Bai2014} models, and thus we assume that they are known a priori. Note that we let $\omega_{ij} = 0$, if $(i,j)$ is in the optimal association $\boldsymbol\tau^*$. Similarly define $\omega_{ik}$ and $\omega_{jk}$. Suppose each AP can serve at most $s_k$ clients as backup association, i.e.,
\begin{align}\label{eq:const-s-k}
  \sum_{i\in\CRm} \left(x_{ik} +  \sum_{j\in\CRp} 2x_{ij}x_{jk} \right)&\leq s_k\,, &&\forall k\in\mathcal A\,.
\end{align}
Therefore, we pose the following optimization problem:
\begin{align}\label{eq:optim-bk}
  \max_{\mathbf x} & \quad\sum_{i\in\mathcal C}\sum_{k\in\mathcal A} \omega_{ik} x_{ik} + \sum_{i\in\CRm}\sum_{j\in\CRp} \sum_{k\in\mathcal A} \omega_{ij} \omega_{jk} x_{ij} x_{jk}
  & \text{s.t.}& \quad\eqref{eq:const-assoc-i}\sim\eqref{eq:const-x},\text{and }\eqref{eq:const-s-k}\,,
\end{align}
which is a variation of optimization problem \eqref{eq:stoch-optim-maxflow}, and the solution approaches for \eqref{eq:stoch-optim-maxflow} could be used for \eqref{eq:optim-bk}. Thus, we mainly focus on optimization problem~\eqref{eq:stoch-optim-maxflow} for the rest of the paper.
\end{enumerate}
}

{Moreover, note that our proposed framework could also capture the coverage issue by letting utility function $f_i$ be the coverage probability, which is a function in SNR. The nature of coverage problem alleviates the necessity of considering the resource sharing at access points, which will substantially simplify the problem. This coverage oriented problem could be solved by similar approaches proposed in this paper, and we leave it for future studies. }

\section{Centralized Solution Approach}
\label{sec:centralized-approach}

In this section, we present the solution approach for optimization problem~\eqref{eq:stoch-optim-maxflow}. The approach is centralized as a first fundamental step to derive a fully distributed solution method later in Section~\ref{sec:distributed-approach}. Specifically, in Section~\ref{subsec:optm-resource}, we propose the optimal rates $\mathbf r^*$ and resource allocation fraction $\mathbf y^*$ given a feasible association $\mathbf x$. Then, using $\mathbf r^*$ and $\mathbf y^*$, in Section~\ref{subsec:multi-assign}, we transform optimization problem~\eqref{eq:stoch-optim-maxflow} into a {multi-dimensional} assignment problem.

\vspace{-0.5cm}
\subsection{Optimal Resource Allocation}
\label{subsec:optm-resource}

Consider random variable $\gamma$ and its estimate $\tilde{\gamma}$. Let $\hat{\gamma}^\eta$ denote the threshold such that
\begin{align}
\label{eq:pr-alpha}
  \Pr (\gamma \leq \hat{\gamma}^\eta| \tilde{\gamma}) = \eta\,.
\end{align}
We assume that the cumulative distribution function~\eqref{eq:cdf-estimate} of the estimates is known. Then, given the estimates of SNR, $\hat{c}^\eta_{ij}$, $\hat{c}^\eta_{ik}$ and $\hat{c}^\eta_{jk}$ can be determined from random variables $c_{ij}$, $c_{ik}$ and $c_{jk}$ by \eqref{eq:pr-alpha}. Then the following lemma provides the optimal rate $\mathbf r^*$ for optimization problem~\eqref{eq:stoch-optim-maxflow}.

\begin{lemma} \label{lem:optimal-rate}
Consider optimization problem~\eqref{eq:stoch-optim-maxflow} and suppose the association variables $\mathbf x$ are fixed and feasible. Then, the optimal rate $\mathbf r^*$ for problem~\eqref{eq:stoch-optim-maxflow} in association $\mathbf x$ is
\begin{enumerate}[(a)]
\item $r_i^* = \hat{c}^\eta_{ik}$, if $x_{ik} = 1$, $\forall i\in\CRm, k\in\mathcal A$,
\item $r_j^* = \hat{c}^\eta_{jk}$, if $\sum_{i\in\mathcal C_-} x_{ij} = 0$ and $x_{jk} = 1$, $\forall j\in\CRp, k\in\mathcal A$,
\item $ r_i^* = \min \left( \hat{c}^\eta_{ij}, \hat{c}^\eta_{jk}/2 \right)$ and $ r_j^* = \hat{c}^\eta_{jk} - r_i^*$, if $x_{ij}x_{jk}=1$, $\forall i\in\CRm, j\in\CRp, k\in\mathcal A$.
\end{enumerate}
\end{lemma}
\begin{IEEEproof}
In the first two cases (a) and (b), client $i$ and relay $j$ associate with AP $k$ directly. It follows that optimal rates for client $i$ and relay $j$ are the maximum achievable rates $\hat{c}^\eta_{ik}$ and $\hat{c}^\eta_{jk}$, respectively, which satisfy constraints~\eqref{eq:const-stoch-linkcap-ij} and \eqref{eq:const-stoch-linkcap-jk}. In the last case (c), client $i$ communicate to AP $k$ via relay~$j$. The optimal rates are the
solutions of the following optimization problem:
{
\begin{subequations}
\label{eq:optim-rate}
\begin{align}
\label{eq:optim-rate-objec}
\max_{r_i,r_j, y_{ik},y_{jk}} & \quad \log (r_i y_{ik}) + \log (r_j y_{jk})\\
\textrm{s.t.} &\quad 0\leq r_i \leq \hat{c}^\eta_{ij}\,, \quad 0\leq r_i + r_j \leq \hat{c}^\eta_{jk}\,,\\
&\quad y_{ik}, y_{jk} \geq 0\,,
\end{align}
\end{subequations}
where \eqref{eq:optim-rate-objec} equals to $\log (r_i r_j) + \log (y_{ik} y_{jk})$. Therefore, the optimal rates for \eqref{eq:optim-rate} are
$ r_i^* = \min \left( \hat{c}^\eta_{ij}, \hat{c}^\eta_{jk}/2 \right)$ and $ r_j^* = \hat{c}^\eta_{jk} - r_i^*$. }It completes the
proof.
\end{IEEEproof}

Before we present the optimal resource allocation strategy for resource allocation fraction $\mathbf y^*$, let us introduce the following useful definition:
\begin{definition}
Denote $n_k$ the number of the clients and the relays that are associated with AP $k$:
\begin{align}
n_k = \sum_{i\in\mathcal C} x_{ik} + \sum_{i\in\CRm} \sum_{j\in\CRp} x_{ij}x_{jk}\,.
\end{align}
\end{definition}

The following lemma establishes the optimal resource allocation at APs.

\begin{lemma} \label{lem:optimal-allocation}
Consider optimization problem~\eqref{eq:stoch-optim-maxflow} and suppose the association variables $\mathbf x$ are fixed and feasible. Then, the optimal resource allocation $\mathbf y^*$ is:{
\begin{enumerate}[(a)]
\item $y_{ik}^* = 1/n_k$, if $x_{ik}=1$, $\forall i\in\mathcal C-_,k\in\mathcal A$,
\item $y_{jk}^* = 1/n_k$, if $\sum_{i\in\mathcal C_-} x_{ij} = 0$ and $x_{jk} = 1$, $\forall j\in\CRp, k\in\mathcal A$,
\item $y_{ik}^* = y_{jk}^* = 2/n_k$, if $x_{ij}x_{jk}=1$, $\forall i\in\CRm, j\in\CRp, k\in\mathcal A$.
\end{enumerate}
}
\end{lemma}
\begin{IEEEproof}
{Since feasible association $\mathbf x$ is given for \eqref{eq:stoch-optim-maxflow}, thus, it is equivalent to find the optimal resource allocation $\mathbf y^*$ for the following optimization problem}
\begin{align}
\label{eq:optim-y}
\max_{\mathbf y} & \quad \sum_{k\in\mathcal A} \left( \sum_{i\in\mathcal C}
 x_{ik} \log y_{ik}  + \sum_{i\in\CRm} \sum_{j\in\CRp} x_{ij} x_{jk} \log y_{ik}  \right)  =   \sum_{k\in\mathcal A} \sum_{i\in \mathcal C_k } \log y_{ik}\,,
\end{align}
{subject to constraints \eqref{eq:const-fract-ij} and \eqref{eq:const-y}.} $\mathcal C_k$ is the set of the clients and the relays that are associated with AP $k$, {for which the corresponding $y_{ik} >0$. Consider the following optimization problem
\begin{subequations}
\label{eq:optim-simple-y}
\begin{align}
\max_{\mathbf y>0} &\quad \prod_{i=1}^l y_i^2 \prod_{i=l+1}^{w} y_i\\
\textrm{s.t.} & \quad \sum_{i=1}^w y_i \leq 1\,,
\end{align}
\end{subequations}
in which $l$ is the number of client-relay pairs that are associated with AP $k$ under given $\mathbf x$, whereas $w$ is the number of connections at AP $k$. That is $l = \sum_{i\in\mathcal C_-}\sum_{j\in\mathcal C_+}x_{ij}x_{jk}$, and $w= \sum_{i\in\mathcal C} x_{ik}$. Moreover, let $y_1, \dots, y_l$ be the resource allocations for the $l$ client-relay pairs. Thus, \eqref{eq:optim-simple-y} captures optimization problem~\eqref{eq:optim-y}. Since \eqref{eq:optim-simple-y} is convex and feasible, by investigating the corresponding KKT conditions, we can find the optimal solution for \eqref{eq:optim-simple-y}, in which $y_1^*=\cdots=y_l^* = 2/n_k$ and $y^*_{l+1}=\cdots=y^*_{w} = 1/n_k$. The calculation is straightforward and therefore is omitted here. }It completes the proof.
\end{IEEEproof}

Based on Lemmas~\ref{lem:optimal-rate} and \ref{lem:optimal-allocation}, {
let $a_{ik}$ and $a_{ijk}$ be} the utility weights that correspond to the association indicator $x_{ik}$ and $x_{ij}x_{jk}$ as:
\begin{enumerate}[(i)]
\item for all client $i \in\mathcal C$ and AP $k\in\mathcal A$,
\begin{align}\label{eq:a_ik}
a_{ik} &= \hat{c}^\eta_{ik}\,,
\end{align}
\item for all client $i\in\CRm$, relay $j\in\CRp$, and AP $k\in\mathcal A$,
\begin{align}\label{eq:a_ijk}
a_{ijk} &= \left\{ \begin{array}{ll} {\hat{c}^\eta_{jk}} & \textrm{if } 2\hat{c}^\eta_{ij} \geq \hat{c}^\eta_{jk}\,,\\
\cfrac{4(\hat{c}^\eta_{jk}-\hat{c}^\eta_{ij})\hat{c}^\eta_{ij}}{\hat{c}^\eta_{jk}} & \textrm{if } 2\hat{c}^\eta_{ij} < \hat{c}^\eta_{jk}\,,
\end{array}\right.
\end{align}
where $\log a_{ijk} = \log (2\bar{r}_i) + \log (2 \bar{r}_j) - \log \tilde{r}_j$, in which $\bar{r}_i$ and $\bar{r}_j$ are the optimal rates if $x_{ij}x_{jk}=1$ (case (c) in
Lemma~\ref{lem:optimal-rate}), whereas $\tilde{r}_j$ is the optimal rate if $x_{jk}=1$ and
$\sum_{i\in\mathcal C_-} x_{ij}= 0$ (case (b) in Lemma~\ref{lem:optimal-rate}).
\end{enumerate}
Then, optimization problem~\eqref{eq:stoch-optim-maxflow} can be rewritten as
\begin{subequations}
\label{eq.intermedia}
\begin{align}
\label{eq.intermedia-obj}
\max_{\mathbf x, \mathbf n} & \quad \sum_{k\in\mathcal A} \left( \sum_{i\in\mathcal C} x_{ik} \log a_{ik}
+ \sum_{i\in\CRm} \sum_{j\in\CRp} x_{ij}x_{jk} \log a_{ijk} \right) - \sum_{k\in\mathcal A} n_k \log n_k\\\label{eq.intermedia-cst}
{\rm s.t.} & \quad n_k = \sum_{i\in\mathcal C} x_{ik} + \sum_{i\in\CRm} \sum_{j\in\CRp}
x_{ij}x_{jk}\,, \quad \forall k\in\mathcal A \\\nonumber
& \quad \eqref{eq:const-assoc-i}, \eqref{eq:const-assoc-j},\text{and } \eqref{eq:const-x}\,,
\end{align}
\end{subequations}
where decision variables are binary indicator $\mathbf x$ and integer variable $\mathbf n = \{n_k|
k\in\mathcal A\}$. Next, we present how to solve this problem. {Note that constraint~\eqref{eq.intermedia-cst} is due to the joint association and relaying mechanism, where a client can have access to an AP either directly or via a relay. Such a constraint is coupled among decision variables, and is not considered in previous works in \cite{ye2013user,Shen2014}. This creates a major technical difficulty, which has never been considered before. Moreover, since equality constraint~\eqref{eq.intermedia-cst} is not affine, problem \eqref{eq.intermedia} is not convex even if binary variable $x_{ij}$ be relaxed to real values~\cite{Boyd2004}. In the following sections, we will develop a novel solution approach for this mixed-integer non-convex optimization problem. }

\subsection{{Multi-dimensional} Assignment Problem}
\label{subsec:multi-assign}

In the section, we transform problem \eqref{eq.intermedia} into a special case of {multi-dimensional} assignment problems. Although the multi-dimensional assignment problems in general have no closed-form solutions~\cite{Bertsekas1998}, they {are convex and }possess useful properties that allow to efficiently compute numerical solutions.

Let introduce virtual client $u$ and relay $v$ that can associate with every relay $j\in\CRp$ and with every client $i\in\CRm$, respectively. Denote ${\CRm}' = \CRm \cup \{u\}$ and ${\CRp}' =
\CRp \cup \{v\}$. { Let introduce binary variable $z_{ijk}\in\{0,1\}$ for all $i\in{\CRm}'$, $j\in{\CRp}'$, $k\in\mathcal A$.
Let $z_{ivk}=x_{ik}$ to imply whether client $i$ is directly associated with AP $k$, whereas let $z_{ujk}=x_{jk}$ to imply whether relay $j$ is associated with AP $k$ without assisting clients. Define $z_{ijk}=x_{ij}x_{jk}$. }
Now define the corresponding utility weight $b_{ijk}$ for all $i\in{\CRm}'$, $j\in{\CRp}'$, and $k\in\mathcal A$
{by letting $b_{ivk}=\log a_{ik}$, $b_{ujk}=\log a_{jk}$, and $b_{ijk} = \log (a_{ijk} a_{jk})$.}
Then, \eqref{eq.intermedia-obj}, \eqref{eq:const-assoc-i}, and \eqref{eq:const-assoc-j} become
\begin{align}
\label{eq.obj-true}
&\sum_{i\in{\CRm}'}
\sum_{j\in{\CRp}'} \sum_{k\in\mathcal A} b_{ijk}z_{ijk} - \sum_{k\in\mathcal A} n_k \log n_k\,,\\\label{eq:const-assoc-z1}
&\sum_{k\in\mathcal A} \sum_{j\in{\CRp}'} z_{ijk} = 1\,, \forall i\in {\color{blue}\CRm}\,,\\\label{eq:const-assoc-z2}
&\sum_{k\in\mathcal A} \sum_{i\in{\CRm}'} z_{ijk} = 1 \,, \forall j \in{\CRp}\,,
\end{align}
respectively, {where the number of clients associated with AP $k$ becomes
\begin{align}
\label{eq:n_k-z}
n_k &= \sum_{i\in{\CRm}'} \sum_{j\in{\CRp}'} w_{ij}z_{ijk}\,, &&\forall k \in\mathcal A\,,
\end{align}
in which constant parameter $w_{ij}=2$, if client $i\in{\mathcal C_-}'$ and relay $j\in{\mathcal C_+}'$, otherwise $w_{ij}=1$.
}
Let $\mathbf z$ denote the binary variables $\{z_{ijk}\,|\,\forall i\in{\CRm}',j\in{\CRp}',k\in\mathcal A\}$. Thus, we are now in position to transform optimization problem~\eqref{eq.intermedia} into a {variant multi-dimensional} assignment problem {with nonlinear objective}:
\begin{subequations}
\label{eq:true-prob}
\begin{align}
\label{eq:true-prob-obj}
\max_{\mathbf z, \mathbf n}  & \quad \sum_{k\in\mathcal A} \sum_{i\in{\CRm}'} \sum_{j\in{\CRp}'}
b_{ijk}z_{ijk} - \sum_{k\in\mathcal A} n_k \log n_k\\
\label{eq.true-prob-cst}
\textrm{s.t.}
&\quad z_{ijk} \in\{0,1\} \,,\quad \forall i \in {\CRm}', j\in{\CRp}', k\in\mathcal A\\
\nonumber
&\quad \eqref{eq:const-assoc-z1},\eqref{eq:const-assoc-z2}, \text{and } \eqref{eq:n_k-z}\,.
\end{align}
\end{subequations}
We remark that the solution to \eqref{eq:true-prob} gives the solution to optimization problem~\eqref{eq.intermedia}. One could solve problem~\eqref{eq:true-prob} by brute force search algorithms. However, the complexity of a brute force algorithm exponentially increases with the number of clients, relays, and APs, making it impractical even for a modest size mmWave network. An alternative solution approach could relax optimization problem~\eqref{eq:true-prob}, in which binary variable $z_{ijk}$ is allowed to take any real value in $[0,1]$. Then, optimization problem \eqref{eq:true-prob} is relaxed to a {non-linear convex optimization problem, which can be solved by centralized methods}, and has the same solution of~\eqref{eq:true-prob} on the condition given by Proposition~\ref{prop.relax-nd} in the following Section~\ref{subsec:primal-dual-decomposition}. However, the centralized approach needs global network information. This requires a centralized coordinator for client association and relaying, which is hard or impossible to have in practice. In the following section, we propose a distributed algorithm to solve \eqref{eq:true-prob} without any central coordinator.

\vspace{-0.5cm}
\section{Distributed Solution Approach}
\label{sec:distributed-approach}
In the previous section, we considered the joint association and relaying optimization problem~\eqref{eq:stoch-optim-maxflow} whose optimal solution returns the transmission rate, the resource allocation, and the association-relay variables. To derive the optimal solution, we developed a solution method by first deriving the solution to the rates (Lemma~\ref{lem:optimal-rate}), then the solution to the resource allocation (Lemma~\ref{lem:optimal-allocation}), and finally we arrived at optimization problem~\eqref{eq:true-prob} whose solution returns the association variables. Such a solution method is centralized, which is impractical in many applications. In this section, we build on these fundamental results and present our distributed solution approach. Specifically, in Section~\ref{subsec:primal-dual-decomposition}, we propose a primal-dual decomposition for optimization problem~\eqref{eq:true-prob} via Lagrangian dual decomposition considering real variable $z_{ijk}$. Then, a primal-dual distributed algorithm is developed in Section~\ref{subsec:primal-dual-algorithm} consisting of a novel distributed auction algorithm. The systematical investigation of the convergence properties of this distributed auction algorithm is presented in Section~\ref{subsec:auction}. Finally, a heuristic load biasing approach is given in Section~\ref{subsec:load-bias} to quicken the convergence. The core results of this section are the derivation of Algorithm~\ref{alg.primal-dual} and its sub-routine Algorithm~\ref{alg_d} to solve optimization problem~\eqref{eq:stoch-optim-maxflow}, and the investigation of the convergence properties of the proposed distributed algorithms.

\vspace{-0.5cm}
\subsection{Lagrangian Dual Decomposition}
\label{subsec:primal-dual-decomposition}

In this section, we develop the first step to arrive at a distributed solution method. In particular, the step consists of a primal-dual decomposition for optimization problem~\eqref{eq:true-prob} via Lagrangian duality considering real variable $z_{ijk}$. We show that the dual problem of \eqref{eq:true-prob} is decoupled into two sub-problems, which can be solved on clients' (and relays') side and AP's side, respectively.

Consider the relaxation of~\eqref{eq:true-prob}, in which binary variable $z_{ijk}$ can take any real value in $[0,1]$. Let $\boldsymbol\lambda=\{\lambda_k|k\in\mathcal A\}$ denote the dual variables corresponding to constraint~\eqref{eq:n_k-z} in the relaxed~\eqref{eq:true-prob}, then the resulting Lagrangian function is
\begin{align*}
\mathcal L (\mathbf z, \mathbf n, \boldsymbol\lambda) =& \sum_{k\in\mathcal A} \sum_{i\in{\CRm}'} \sum_{j\in{\CRp}'}
\left( b_{ijk}\!-\!\lambda_k w_{ij} \right)z_{ijk} \!+\! \sum_{k\in\mathcal A} n_k \left(\lambda_k - \log n_k \right).
\end{align*}
Thus, the dual problem of the relaxed problem~\eqref{eq:true-prob} is
\begin{align}
\label{eq.dual-main}
\min_{\boldsymbol\lambda} \quad g_n (\boldsymbol\lambda) + g_z (\boldsymbol\lambda)\,,
\end{align}
where $g_n(\boldsymbol\lambda)$ and $g_z(\boldsymbol\lambda)$ are
\begin{align}
\label{eq.sub-AP}
g_n(\boldsymbol\lambda)\! &=\! \max_{\mathbf n\geq 0} \quad \sum_{k\in\mathcal A} \,n_k \left(\lambda_k - \log n_k \right)
\end{align}
\begin{align}
\label{eq.sub-multiassign}
g_z (\boldsymbol\lambda)\! &=\! \left\{ \begin{array}{lll} \!\max\limits_{\mathbf z} \!& \! \sum\limits_{k\in\mathcal A}
\sum\limits_{i\in{\CRm}'} \sum\limits_{j\in{\CRp}'} \left( b_{ijk}\!-\!\lambda_k w_{ij} \right)z_{ijk}\\
\!\textrm{s.t.}\! & \! z_{ijk} \!\in \![0,1]\,,  \forall i\!\in\!{\CRm}', j\!\in\!{\CRp}', k\!\in\!\mathcal A\\
& \! \eqref{eq:const-assoc-z1} \text{ and }\eqref{eq:const-assoc-z2}\,.
\end{array} \right.
\end{align}

Now we are in the position to give the condition for which the relaxation does not give a suboptimal solution.

\begin{proposition}
\label{prop.relax-nd}
Denote by $\mathbf n^* (\boldsymbol\lambda^*)$ and $\mathbf z^*(\boldsymbol\lambda^*)$ the maximizers of subproblems \eqref{eq.sub-AP} and \eqref{eq.sub-multiassign}, respectively, where $\boldsymbol \lambda^*$ the optimal solution to dual problem~\eqref{eq.dual-main}. If maximizer $\mathbf z^*(\boldsymbol\lambda^*)$ of \eqref{eq.sub-multiassign} is unique, then $\mathbf n^* (\boldsymbol\lambda^*)$ and $\mathbf z^*(\boldsymbol\lambda^*)$ are the optimal solutions for optimization problem~\eqref{eq:true-prob}.
\end{proposition}

\begin{IEEEproof}
Denote by $\mathcal R$\eqref{eq:true-prob} the relaxation of optimization problem~\eqref{eq:true-prob}, in which binary variables $z_{ijk}$ can take any value in $[0,1]$. Note that optimization problem~\eqref{eq.dual-main} is the dual problem of $\mathcal R$\eqref{eq:true-prob}. Since the constraints in $\mathcal R$\eqref{eq:true-prob} are all linear equalities, thus the Slater condition reduces to feasibility~\cite{Boyd2004}. Optimization problem~$\mathcal R$\eqref{eq:true-prob} is feasible, because there always exists feasible solutions by letting $z_{ivk} = 1$ and $z_{ujk} = 1$ for all client $i$, relay $j$ and AP $k$. Thus, strong duality holds, that is, the optimal values of $\mathcal R$\eqref{eq:true-prob} and \eqref{eq.dual-main} are the same. Therefore, the optimal solutions of primal problem $\mathcal R$\eqref{eq:true-prob} can be obtained by solving dual problem~\eqref{eq.dual-main}. That is, $\mathbf z^*(\boldsymbol\lambda^*)$ and $\mathbf n^* (\boldsymbol\lambda^*)$  are the optimal solutions of primal problem $\mathcal R$\eqref{eq:true-prob}. Moreover, given that we are considering a linear programming problem, there exist optimal solutions for assignment problem~\eqref{eq.sub-multiassign}, which are either 0 or 1~\cite{Bertsekas1998}.\footnote{\color{blue}In Section~\ref{subsec:auction}, we show that linear optimization~\eqref{eq.sub-multiassign} can be transformed into an asymmetric assignment problem.} Therefore, if maximizer $\mathbf z^*(\boldsymbol\lambda^*)$ of \eqref{eq.sub-multiassign} is unique, it must be binary, which indicates that $\mathbf z^*(\boldsymbol\lambda^*)$ and $\mathbf n^*(\boldsymbol \lambda^*)$ are feasible for optimization problem~\eqref{eq:true-prob}, which completes the proof.
\end{IEEEproof}
The previous proposition suggests that the relaxation of the binary constraints, with the dual problem~\eqref{eq.dual-main}, can be useful to derive the solution to problem~\eqref{eq:true-prob}. Specifically, if the condition given by the previous proposition is satisfied, then these two optimization problems have exactly the same solution. Instead, if the condition given in the previous proposition is not fulfilled, we still can find sub-optimal solutions by solving dual problem~\eqref{eq.dual-main}. The sub-optimal solution is very close to the optimal one, as studied via extensive numerical simulations. In the next subsection, we develop a solution algorithm for dual problem~\eqref{eq.dual-main}.

\vspace{-0.5cm}
\subsection{The Distributed Algorithm}
\label{subsec:primal-dual-algorithm}

\begin{subalgorithms}
\label{alg.primal-dual}
\begin{algorithm}[t]
\caption{Distributed Algorithm for Clients and Relays} \label{alg.primal-dual-client}
\footnotesize
\begin{algorithmic}[1]
\State Measure the SNR by using pilot signals, and receive $\lambda_k$ broadcast by each AP %
\State Solve~\eqref{eq.sub-multiassign} by the distributed auction algorithm summarized in Algorithm~\ref{alg_d} (described in
Section~\ref{subsec:auction}) %
\end{algorithmic}
\end{algorithm}

\begin{algorithm}[t]
\caption{Distributed Algorithm for AP $k$} \label{alg.primal-dual-AP}
\footnotesize
\begin{algorithmic}[1]
\State Solve \eqref{eq.sub-AP} by setting $n_k (t+1) = \exp \left( \lambda_k (t)-1 \right)$ %
\State Update $\lambda_k$ by letting
\begin{align}
\label{eq:lambda-update}
\lambda_k (t\!+\!1) = \lambda_k (t) - \delta (t) \bigg( n_k (t)\! -\!
\sum_{i\in\CRm}\! \sum_{j\in\CRp} w_{ij} z_{ijk} \bigg)\,.
\end{align}
\end{algorithmic}
\end{algorithm}
\end{subalgorithms}

Optimization problem \eqref{eq.dual-main} is convex and can be solved in a distributed manner. To this end, we propose Algorithm~\ref{alg.primal-dual}. In the algorithm, the clients measure the SNR by using pilot signals from all relays and APs, while each AP $k$ initializes and
broadcasts the price of each AP $k$, $\lambda_k(t)$ at each discrete time $t$. This price is determined by the load situation as stated in \eqref{eq:lambda-update}. After receiving $\lambda_k(t)$, the clients and the relays solve multi-dimensional assignment
problem~\eqref{eq.sub-multiassign} by the distributed auction algorithm, which will be presented in
Section~\ref{subsec:auction}. These steps are shown in Algorithm~\ref{alg.primal-dual-client}. Then, each AP $k$ updates $\lambda_k (t+1)$ by the gradient method, and broadcasts $\lambda_k (t+1)$ to the clients, as shown in Algorithm~\ref{alg.primal-dual-AP}. Note that step size $\delta(t)$ is properly chosen according to \cite{Boyd2004} to guarantee the convergence. {We remark here that in each iteration of Algorithm~\ref{alg.primal-dual}, only the current values of dual variable $\boldsymbol\lambda$ are required for the clients.}
\vspace{-0.5cm}
\subsection{Distributed Auction Algorithm}
\label{subsec:auction}

{Contrary to the works in \cite{ye2013user,Shen2014}, it is not trivial to find the optimal association for clients given dual variable $\boldsymbol\lambda$, since sub-optimization problem~\eqref{eq.sub-multiassign} (with binary variables) is a multi-dimensional assignment problem, which is in general NP hard to solve~\cite{Bertsekas1998}. }In this section, {sub-optimization} problem~\eqref{eq.sub-multiassign} is first transformed into an asymmetric assignment problem. Then, we propose a novel distributed solution approach based on the auction algorithm~\cite{Bertsekas1998}. Given $\boldsymbol\lambda$ by~\eqref{eq:lambda-update}, consider the following derivations
\begin{align} \label{eq.b2beta}
\tilde{k}_{ij} &= \argmax_{k\in\mathcal A} \, \left( b_{ijk} - \lambda_k w_{ij}\right)\,, &
\alpha_{ij} &= b_{ij\tilde{k}_{ij}} - \lambda_{\tilde{k}_{ij}} w_{ij}\,, &
\tilde{z}_{ij} &= z_{ij\tilde{k}_{ij}}\,,
\end{align}
and let $\alpha_{uv} = 0$. Those derivations are to associate the relays with their best APs. Note that AP $\tilde{k}_{iv}$, obtained by \eqref{eq.b2beta}, is the best AP for client $i$ without assistance from relays. Therefore, problem~\eqref{eq.sub-multiassign} can be transformed to a {variant} assignment problem~\cite{Bertsekas1998}:
\begin{subequations}
\label{eq.sub-assign}
\begin{align}
\max_{\tilde{\mathbf z}} &\quad \sum_{i\in{\CRm}'} \sum_{j\in{\CRp}'} \alpha_{ij} \tilde{z}_{ij} \\
\textrm{s.t.} & \quad \sum_{j{\in{\CRp}'}} \tilde{z}_{ij} = 1\,, & &\forall i \in {\CRm} \\
& \quad \sum_{i\in{\mathcal C_-}'} \tilde{z}_{ij}  = 1\,, & & \forall j \in {\CRp}\\
& \quad \tilde{z}_{ij} \geq 0\,, && \forall i \in{\CRm}', j\in{\CRp}'
\end{align}
\end{subequations}
where $\tilde{\mathbf z} = \{\tilde{z}_{ij}| i \in{\CRm}', j\in{\CRp}' \}$ is the decision variable. The solution of \eqref{eq.sub-multiassign} can be obtained from \eqref{eq.sub-assign} and \eqref{eq.b2beta}. Note that there may be multiple optimal solutions for \eqref{eq.sub-multiassign}, since clients may achieve the same utility with different associations.

The classic centralized auction algorithm established in~\cite{Bertsekas1998} {needs sufficient extension before it could be} used to solve~\eqref{eq.sub-assign}, {not only because two entities (virtual client $u$ and relay $v$) could be associated with multiple other entities (relays and clients respectively), but also because it has the drawback of needing a central coordinator to manage the prices and the bids}. In the following, we develop a novel distributed auction algorithm, Algorithm~\ref{alg_d}, for assignment problem~\eqref{eq.sub-assign}. Algorithm~\ref{alg_d} is
given by the application of Algorithm~\ref{alg_d_c} by the clients, and Algorithm~\ref{alg_d_r} by the
relays, as we describe in the following.

In Algorithm~\ref{alg_d}, the vector $P_i \in \mathds{R}^{N}$ denotes the prices vector for the relays
(stored in client $i$), where $N$ is the cardinality of ${\CRp}$. Let $p_j$ denote the price of a relay $j$ (stored in relay $j$), and virtual relay $v$ represents the best AP $\tilde{k}_{iv}$ obtained from \eqref{eq.b2beta} for client $i$. In what follows, we present the basic steps of Algorithm~\ref{alg_d}.

Initially, we set the prices of all the relays to zero, and choose desired value for the design parameter $\epsilon$ (we show below that it can be chosen to ensure a desired optimality). On the client side (Algorithm~\ref{alg_d_c}), every client $i$ fulfilling the condition in Line 8 finds the best relay $j_i$ using the local knowledge of the prices. In Lines 9$\sim$12, client $i$ calculates the largest bid for the relay $j_i$. Then, it sends the request to $j_i$. On the relay side (Algorithm~\ref{alg_d_r}), when relay $j_i$ receives the request from clients with different bids, it chooses the best client $i_j$ that provides the highest bid and higher price compared to the old price $p_j$. Relay $j_i$ updates its price and feedbacks the latest price to the clients, as described in Lines 4$\sim$6 and Line 8. The auction algorithms terminate when there are no client requests. {We remark here that in each iteration, the bids $\beta$ to the relays, the prices $p$ of relays, and the decision (\textbf{yes} or \textbf{no}) are exchanged in the mmWave networks.}

\begin{subalgorithms}
\label{alg_d}
\begin{algorithm}[t]
\caption{Distributed Auction Algorithm for Client $i$} \label{alg_d_c}
\footnotesize
\begin{algorithmic}[1]
\State Initialize $j_i=v, P_i=\mathbf{0}$ %
\While {\textbf {true}} \If {receive {\bf no} and new price $p_{j_i}$ from $j_i$} %
\State Disconnect to relay $j_i$, connect to AP $k_i^*$, $[P_i]_{j_i} \gets p_{j_i}$, and $j_i \gets k_i^*$ %
\EndIf %
\If {$j_i = v \neq \argmax\limits_{j\in {\CRp}'} \left\{ \alpha_{ij} - [P_i]_j \right\}$ } %
\State $j_i\rq{} \gets \argmax\limits_{j\in {\CRp}'} \left\{ \alpha_{ij} - [P_i]_j \right\}$, $\theta_i \gets \max\limits_{j\in {\CRp}'} \left\{\alpha_{ij} - [P_i]_j \right\}$, $\omega_i \gets \max\limits_{j\in {\CRp}', j\neq j_i} \left\{ \alpha_{ij}- [P_i]_j \right\}$, and $\beta_{i{j_i}} \gets p_{j_i}+ \theta_i- \omega_i + \epsilon$ %
\State Send request with $\beta_{i{j_i}}$ to relay $j_i\rq{}$, receive respond, (\textbf{yes} or \textbf{no}) and $p'_{j_i}$ 
\If{respond contains \textbf{yes}} %
\State Connect to object $j_i\rq{}$, and $j_i\gets j_i\rq{}$ %
\EndIf %
\State $p_{j_i} \gets p'_{j_i}$ \EndIf \EndWhile
\end{algorithmic}
\end{algorithm}

\begin{algorithm}[t]
\caption{Distributed Auction Algorithm for Relay $j$} \label{alg_d_r}
\footnotesize
\begin{algorithmic}[1]
\State Initialize the client $i_j=\emptyset$, and price $p_j=\alpha_{uj}$ %
\If {receive request from clients $i$ and $\beta_i$} 
\If{$\beta_i - p_j \geq \epsilon$} %
\State Send \textbf{yes} and $p_j$, to client $i$, send \textbf{no} and $p_j$, to client $i_j$, and connect to client $i$, and $i_j \gets i$, $p_j \gets \beta_i$ %
\Else %
\State Send \textbf{no} and $p_j$, to client $i$ %
\EndIf \EndIf
\end{algorithmic}
\end{algorithm}
\end{subalgorithms}

\begin{proposition}
\label{prop.finity-convergence}
Let $\epsilon$ be a desired positive constant. Denote by $M$, $N$ and $K$ the cardinalities of sets ${\CRm}$, ${\CRp}$ and $\mathcal A$, respectively. Algorithm~\ref{alg_d} terminates within $M N^2 \lceil \Delta /\epsilon \rceil$ iterations, where $\Delta = \max \alpha_{ij} - \min \alpha_{ij}$.
\end{proposition}

\begin{IEEEproof}
In every iteration, there may exist clients that are not yet informed of the latest price of some relays.
This implies that these clients may place low bids for expensive relays. However, these clients will be
informed after the biding. Based on this observation, we can disregard all lower bids and only consider
bids by informed clients that increase the actual price of the relays. Hence, we only need to show that
every relay can only receive a finite number of such bids.

Whenever $l$ bids are placed for a relay, its price must increase by at least $\epsilon l$. Thus, when $l$
is sufficiently large, this relay will become too expensive to be attractive compared to other relays that
have not yet received any bids. It follows that there is a limited number of bids that any relay can
receive by informed clients. Therefore, the auction will continue until each one of the clients has been
associated with one relay.

In the worst case, we consider that all the clients persistently place minimum bid increments $\epsilon$.
Furthermore, they will not win the object until the local price vectors in clients is updated to the latest. Without considering the price update, the number of iterations of the auction algorithm is bound by $N \lceil \Delta/\epsilon \rceil$, because every client $i$ will eventually be associated with the
best AP when the benefit of all the relay nodes in $\CRp$ is lower than that of the best AP.
Meanwhile, in every iteration, the throughput benefit decreases monotonically at least by $\epsilon$. On
the other side, the number of iterations for price update is bounded by $MN$. Thus the number of iterations of the distributed algorithms is bounded by $M N^2 \lceil \Delta/\epsilon \rceil$, which completes the
proof.
\end{IEEEproof}

\begin{remark}
Note that the previous bound is conservative. This bound is based on the absence of broadcast transmissions in the network. If every relay $j$ can broadcast its latest prices to clients, then we can show that the iterations are bounded by $N^2 \lceil \Delta/\epsilon \rceil$.
\end{remark}

\begin{proposition}
\label{prop.convergence-distributed}
Let $\epsilon$ be a desired positive constant. The final assignment obtained by Algorithm~\ref{alg_d} is within $M\epsilon$ of the optimal assignment benefit of problem~\eqref{eq.sub-assign}. The final assignment is optimal if $\alpha_{ij}$, $\forall i\in{\CRm}, j\in{\CRp}$, is integer, and $\epsilon < 1/M$.\footnote{If all benefits are rational numbers, they can be scaled up to integer by multiplication with a suitable common number.}
\end{proposition}

In order to prove this proposition, we need some technical intermediate results. Consider the dual
problem of \eqref{eq.sub-assign}:
\begin{subequations}
\label{eq.dual-assign}
\begin{align}
\label{eq.dual-assign-obj}
\min_{p_j,\pi_i} & \quad \sum_{i\in\CRm} \pi_i + \sum_{j\in\CRp} p_j \\
\label{eq.dual-assign-cst-1}
\textrm{s.t.} & \quad \pi_i + p_j \geq \alpha_{ij}\,, && \forall i\in\CRm, j \in \CRp\\
\label{eq.dual-assign-cst-2}
&\quad \pi_i \geq \alpha_{iv}\,, &&\forall i \in\CRm\\
\label{eq.dual-assign-cst-3}
&\quad p_j \geq \alpha_{uj}\,, && \forall j \in\CRp
\end{align}
\end{subequations}
where $\pi_i$ is the Lagrangian multiplier introduced to represent the benefit of each client $i$, $p_j$
represents the price for relay $j$. Moreover, let $p_v = 0$ for the virtual relay $v$. Furthermore, we give the {novel} definition of $\epsilon$-Complementary Slackness ($\epsilon$-CS) inspired by~\cite{Bertsekas1998}:

\begin{definition} \label{def.eps-CS}

Let $\epsilon>0$ be a fixed scalar. An association $\mathcal S$ and a price vector $p$ satisfy
$\epsilon$-CS if
\begin{align*}
\alpha_{ij} - p_{j_i} &\geq \max_{j\in\CRp} \{ \alpha_{ij} - p_j\} - \epsilon\,,
&& \forall (i,j_i) \in\mathcal S\,,\\
p_j & \geq \alpha_{uj}\,, && \forall j \in \CRp\,.
\end{align*}
\end{definition}
Thus, the following proposition clarifies the significance of the preceding $\epsilon$-CS condition.

\begin{proposition} \label{prop.CS-Mepsilon}
If a feasible association $\mathcal S$ satisfies the $\epsilon$-CS conditions together with a vector $p$,
then $\mathcal S$ is within $M\epsilon$ of being optimal for the optimization
problem~\eqref{eq.sub-assign}.
\end{proposition}
\begin{IEEEproof}
Denote by $A^*$ and $D^*$ the optimal objective value for primal problem \eqref{eq.sub-assign} and dual
problem \eqref{eq.dual-assign}, respectively. From the strong duality theorem, we have $A^* \leq D^*$.

Consider any feasible association $\mathcal S= \{(i,j_i)\}$ together with $p$ satisfying $\epsilon$-CS
conditions. Moreover, let $\pi_i = \alpha_{ij_i} - p_{j_i}$ for all $i$, thus since $p_v=0$, we have
\begin{align*}
\pi_i + p_j &\geq  \alpha_{ij} - \epsilon\,, && \forall i\in\CRm,j\in\CRp\,,\\
\pi_i & \geq \alpha_{iv} - \epsilon\,, && \forall i\in\CRm\,.
\end{align*}
Now let $\hat{\pi}_i = \pi_i + \epsilon$ for all $i$, which together with $p$ is feasible for the dual
problem, and satisfies
\begin{align*}
\hat{\pi}_i + p_{j_i} &= \alpha_{ij_i} + \epsilon\,,\quad&&\forall (i,j_i) \in\mathcal S, i\neq u, j\neq v\\
\hat{\pi}_i & = \alpha_{iv} + \epsilon \,. \quad && \forall (i,v) \in\mathcal S
\end{align*}
Denote by ${\CRm}^v$ the set of clients associated with virtual relay $v$, and by ${\CRp}^u$ the set of relays associated with virtual client $u$. Thus, we have
\begin{align*}
A^* & \geq \sum_{(i,j_i)\in\mathcal S}\! \alpha_{ij_i} = \sum_{i\in\CRm \setminus {\CRm}^v}\! \alpha_{ij_i} \! +\! \sum_{i\in{\CRm}^v}\! \alpha_{iv} \! +\! \sum_{j_i\in{\CRp}^u} \!\alpha_{uj_i}\\
& = \sum_{i\in\CRm \setminus {\CRm}^v}\! \hat{\pi}_i \! +\! \sum_{j_i\in\CRp \setminus {\CRp}^u}\!
p_{j_i} \! +\! \sum_{i\in{\CRm}^v} \!\hat{\pi}_i \! +\! \sum_{j_i\in{\CRp}^u}\! p_{j_i} \!-\! M\epsilon\\
& = \sum_{i\in\CRm} \hat{\pi}_i + \sum_{j\in\CRp} p_j - M\epsilon \geq D^* - M\epsilon\,,
\end{align*}
which completes the proof.
\end{IEEEproof}

Now, we are ready to prove Proposition~\ref{prop.convergence-distributed}.
\begin{IEEEproof}
Based on Proposition~\ref{prop.CS-Mepsilon}, the result will follow once we prove that
Algorithm~\ref{alg_d} preserves $\epsilon$-CS conditions upon termination of the algorithm.

Assume that $\epsilon$-CS conditions are satisfied at the start of an iteration. Let $(\pi,p)$ and
$(\bar{\pi},\bar{p})$ be the benefit-price pair before and after the iteration, respectively. Suppose that
relay $j^*\in\CRp$ receives a bid from client $i\in\mathcal C_-$ and is assigned to $i$ during the
iteration. Then we have
\begin{align}
\label{eq:pj-star}
\bar{p}_{j^*} = \alpha_{ij^*} - \max_{j\in{\CRp}', j\neq j^*} \{\alpha_{ij} - [P_i]_{j}\} + \epsilon\,.
\end{align}
Since $\bar{p}_j \geq p_j \geq [P_i]_j $ for all $j\in\CRp$, then \eqref{eq:pj-star} implies
\begin{align*}
\alpha_{ij^*} - \bar{p}_{j^*} &= \max_{j\in{\CRp}', j\neq j^*} \{\alpha_{ij} - [P_i]_{j}\} - \epsilon \geq \max_{j\in{\CRp}'} \{\alpha_{ij} - \bar{p}_{j}\} - \epsilon\,,
\end{align*}
which shows that the $\epsilon$-CS condition continues to hold after the assignment phase of an iteration
for all pairs $(i,j^*)$ that entered the assignment during the iteration.

Consider also any pair $(i,j^*)$ that belonged to the assignment just before the iteration, and also
belongs to the association after the iteration. Then $j^*$ must not have received a bid during the
iteration, so $\bar{p_{j^*}} = p_{j^*}$. Therefore, the $\epsilon$-CS conditions hold for all $(i,j^*)$
that belong to the association after the iteration. It completes the proof.
\end{IEEEproof}

{Propositions~\ref{prop.finity-convergence} and \ref{prop.convergence-distributed} show that the proposed distributed auction algorithm provides the sub-optimal solutions in finite iterations without any central coordinator for a variant assignment problem~\eqref{eq.sub-assign}, and equivalently for a special multi-dimensional assignment problem~\eqref{eq.sub-multiassign}.}

\vspace{-0.5cm}
\subsection{Load Biasing}\label{subsec:load-bias}

Algorithm~\ref{alg.primal-dual}, proposed in Section~\ref{subsec:primal-dual-algorithm}, needs a few iterations to converge, which may be time consuming. In this subsection, we introduce a simple approach, namely load biasing, which provides near optimal performance compared to Algorithm~\ref{alg.primal-dual}. The load biasing scheme is inspired by the range expansion~\cite{ye2013user}. In the load biasing scheme, each AP $k$ broadcasts biasing factor $\mu_k$ to all the available clients and relays, instead of the dual variable $\lambda_k$. According to Algorithm~\ref{alg.primal-dual}, the optimal $\mu_k^*$ equals to $\lambda_k^*$, namely it is the optimal solution $\lambda_k^* = \log n_k^* + 1$. The load biasing consists in estimating $\mu_k^*$ in advance. After running Algorithm~\ref{alg.primal-dual} a number of iterations, then a good estimation is given by the empirical mean, i.e., we let $\hat{\mu}_k = \log \bar{n}_k + 1$, where $\bar{n}_k$ is the average number of clients associated with AP $k$.

\vspace{-0.5cm}
\section{Numerical Results}
\label{sec:numerical-examples}

In this section, we numerically evaluate the performance of the proposed solution algorithms for optimization problem~\eqref{eq:stoch-optim-maxflow}, namely Algorithm~\ref{alg.primal-dual} and its sub-routine Algorithm~\ref{alg_d}.

\begin{figure}[t!]
\centering
\subfloat[]{
\includegraphics[width = 0.4\columnwidth]{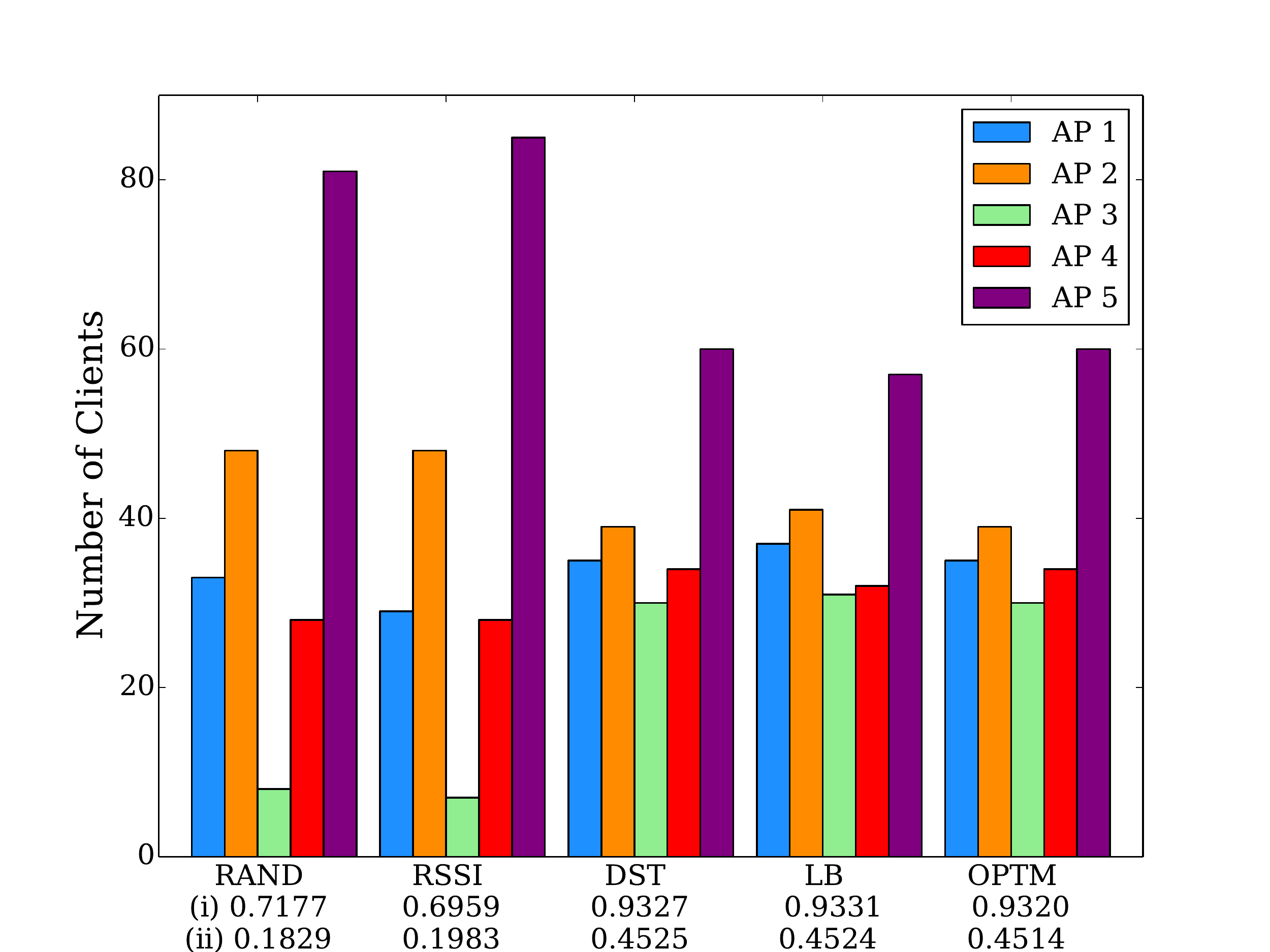}
\label{fig.loads}}
\subfloat[]{
\includegraphics[width = 0.48\columnwidth]{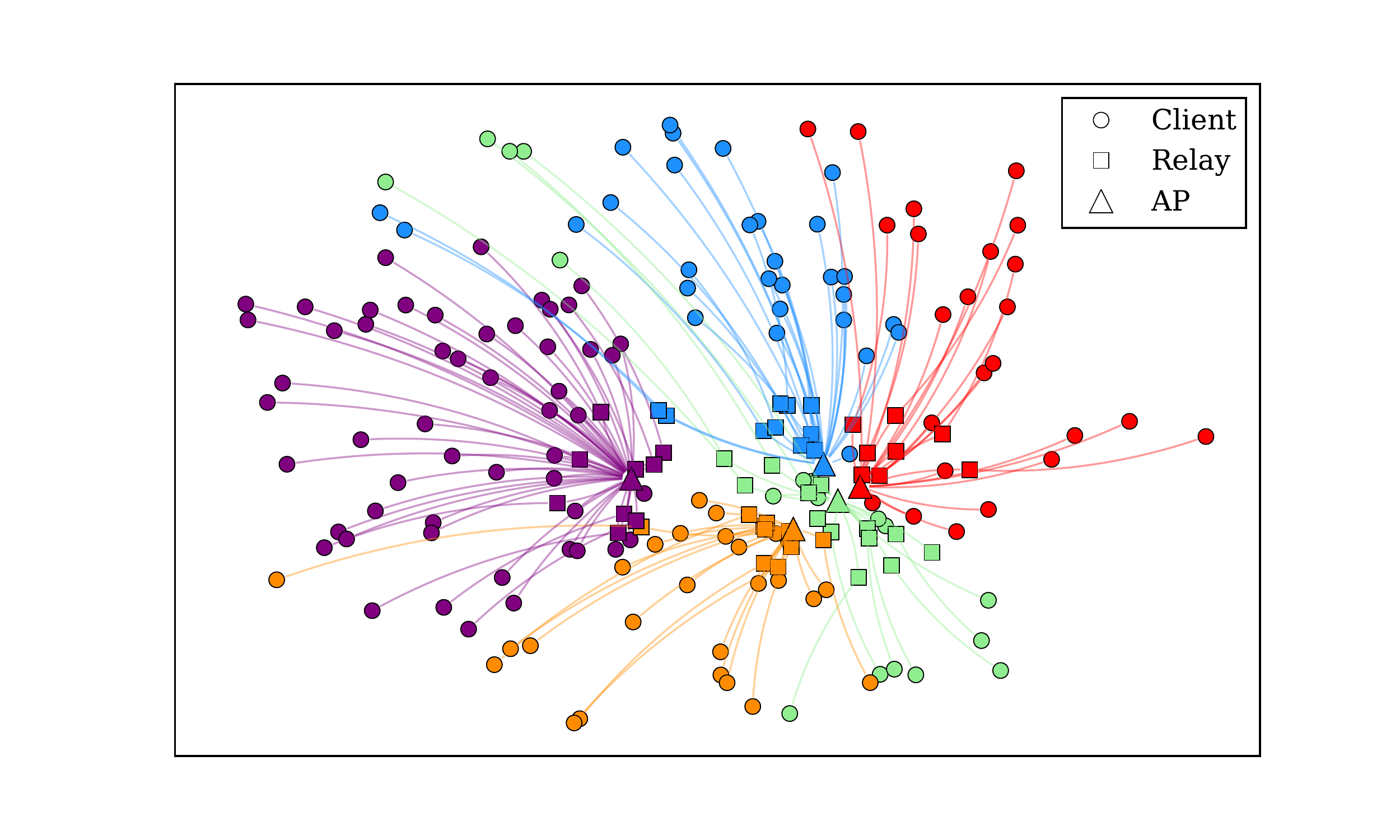}
\label{fig.illustration of mmW}
}
\caption{(a) Illustration of the average number of clients per AP, where APs 1$\sim$5 correspond to different AP in the mmWave network. The values beneath the algorithms' names are the corresponding Jain's fairness indexes: (i) is association fairness, (ii) is throughput fairness. (b) Illustration of a sample solution of DST in Fig.~\ref{fig.loads}. }
\label{fig.rate}
\vspace{-0.8cm}
\end{figure}

We consider a mmWave network operating at 60~GHz, with 150 clients, 50 relays, and 5 APs. {The narrowband geometrical channel model with one path for every transmitter-receiver pair is used. The channel model between transmitter $i$ and receiver $j$ contains a zero mean complex normal random variable with variance $10^{-0.1 L_{ij}}$, where $L_{ij}$ is the corresponding path loss and consists of a constant attenuation, a distance dependent attenuation, and a large scale Lognormal fading. Parameters of the channel model depend on being in LoS or in NLoS and are adopted from \cite[Table~I]{Akdeniz2014}. }In the first set of experiments, we fix the locations of APs {(distributed not uniformly as illustrated in Fig.~\ref{fig.illustration of mmW})}, whereas the clients and relays are uniformly distributed at random over this area. The main parameters used in simulations are listed in Table~\ref{tab_sim_para}. To evaluate the performance of Algorithms~\ref{alg.primal-dual} and~\ref{alg_d}, we run Monte-Carlo simulations over $100$ experiments. We run at most $200$ and $500$ iterations in Algorithm~\ref{alg.primal-dual} and in Algorithm~\ref{alg_d}, respectively. In the following, we use $\epsilon = 0.01$ and assume perfect estimates of the achievable rates (link capacities) when not stated in the description. For benchmarking purpose, we consider the following algorithms:
\begin{inparaenum}
  \item RAND: Random association policy;
  \item RSSI: RSSI-based association policy~\cite{802_15_3c};
  \item DST: Our proposed distributed algorithm in Section~\ref{subsec:primal-dual-algorithm} given in Algorithm~\ref{alg.primal-dual} and its sub-routine Algorithm~\ref{alg_d};
  \item DST-R: DST without relaying;
  \item LB: DST modified by our proposed load biasing policy in Section~\ref{subsec:load-bias};
  \item OPTM: The optimal association policy, which is the solution of the optimization problem~\eqref{eq:stoch-optim-maxflow} obtained by using a centralized binary integer programming solver \texttt{intlinprog} in Matlab.
\end{inparaenum}
Furthermore, we compute the well known Jain's fairness index~\cite{jain1998quantitative} for every realization of every scenario, and find its average. To evaluate the load balancing performance, we consider two definitions for fairness:
\begin{inparaenum}
\item association fairness: the closeness of the number of clients associated to different APs,
\item throughput fairness: the closeness of the throughput of clients.
\end{inparaenum}
For both definitions, we evaluate Jain's fairness index.

\begin{table}[t]
  \centering
  \caption{Simulation Parameters for mmWave Networks}\label{tab_sim_para}
  \scriptsize
  \begin{tabular}{ll}
    \toprule
    Parameters & Value \\
    \midrule
    system bandwidth    & 2.16 GHz \\
    path loss exponent  & 2.5 \\
    operating frequency         & 60 GHz \\
    noise power spectral density & -174 dBm/Hz \\
    noise figure of the receiver & 6 dB\\
    \bottomrule
  \end{tabular}
\end{table}

We analyze the performance of our proposed association and the existing association policies from the literature in terms of network throughput and fairness.

Fig.~\ref{fig.loads} illustrates the load distribution among different APs and the average Jain's fairness index for different association and relaying algorithms, reported beneath the corresponding algorithm's name. The RAND and the RSSI associations result in very unbalanced loads. In particular, AP~5 serves around the half of the clients, whereas AP~3
serves far fewer than 10 clients in both schemes. This unbalanced number of clients per AP, besides resulting in a poor network throughput performance, devastates the association fairness. Moreover, our proposed optimization problem~\eqref{eq:stoch-optim-maxflow} substantially improves the throughput fairness compared to RAND and RSSI-based associations. Note that at every AP, we have equal allocation to maximize network-wide proportional fairness, as shown in Lemma~\ref{lem:optimal-allocation}. It indeed means that a closer client to the AP receives the same share of resources as a very far away one, so we do not have fairness in rate allocation inside every AP, imposed by the objective function~\eqref{eq.objective-log}. The DST provides almost the same load balancing performance as that obtained by the OPTM, which is consistent to the theoretical analysis in Section~\ref{subsec:auction}. Furthermore, the LB provides near optimal load balancing with lower complexity compared to the DST. Adding more APs increases the degree of freedom we have for load balancing and throughput enhancement in the network. While the DST, the LB, and the OPTM solutions optimally leverage these extra degree of freedoms, the performance enhancement of the RAND and the RSSI approaches are not similar to load-aware association and resource allocation policies.

Fig.~\ref{fig.illustration of mmW} shows one sample association of the scenarios used in Fig.~\ref{fig.loads}, and confirms loose meaning of the regular-shape non-overlapping serving areas of the APs in mmWave networks. The clients prefer to be served by farther but less-loaded APs, instead of having higher peak rate per channel use but lower number of the channel uses (being served by a closer but over-loaded AP), as it is clear from the green clients in Fig.~\ref{fig.illustration of mmW}. In fact, the green AP (roughly) prefers to serve closer clients, and relays connect farther clients to the less-loaded APs.



\begin{figure}[t!]
\centering
\subfloat[\footnotesize CDF for the rates]{
\includegraphics[width=0.4\columnwidth]{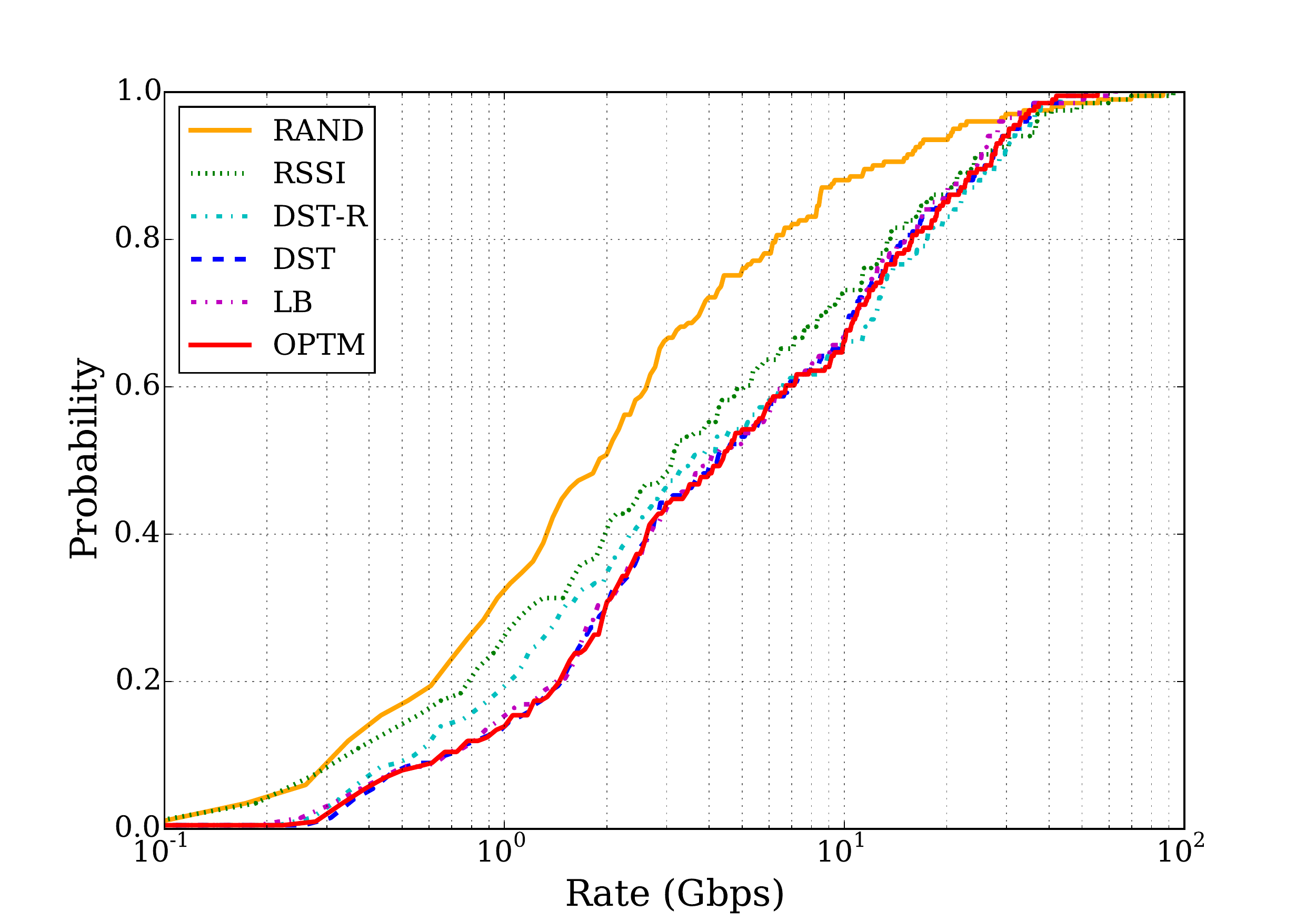}
\label{fig.rate-cdf}}
\subfloat[\footnotesize Rate gain vs. the probability]{
\includegraphics[width=0.4\columnwidth]{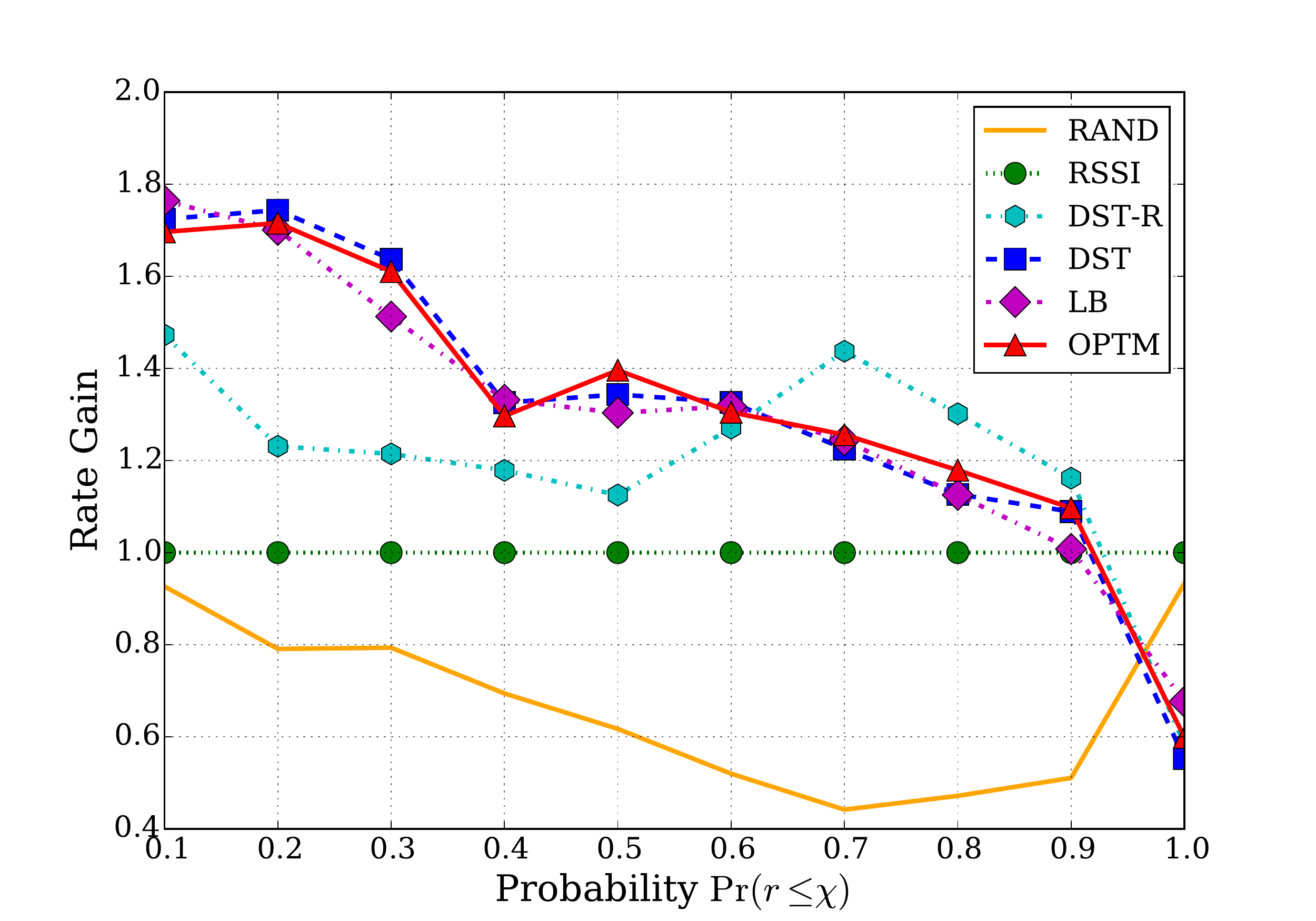}
\label{fig.rate-ratio}
}
\caption{Comparison of the rates of different association approaches: (a) the cumulative distribution functions of the rates, (b) the radio of rate $\chi$ versus the probability of the rates less than $\chi$.}
\label{fig.rate}
\end{figure}

Fig.~\ref{fig.rate} compares achievable rate of different association and resource allocation policies. Specifically, Fig.~\ref{fig.rate-cdf} illustrates the cumulative distribution functions (CDF) for the rates. DST and LB substantially decrease the number of clients with low rates compared to the RAND and the RSSI policies, which provides proportional fairness among the clients. Fig.~\ref{fig.rate-ratio} depicts the gain of rate $\chi$ versus the probability $\Pr (r\leq \chi)$ for the OPTM, the DST, the LB and the RAND approaches compared to the RSSI. As can be observed that OPTM, DST and LB provide 30\%\:$\sim$\:80\% higher rates than RSSI, and 20\%\:$\sim$\:50\% higher rates than DST-R at lower rates. Furthermore, in both Figs.~\ref{fig.rate-cdf} and \ref{fig.rate-ratio}, the outputs of the OPTM and the DST almost overlap, whereas the LB provides a near optimal performance.

\begin{figure}[t!]
\centering
\subfloat[\footnotesize All clients and relays]{
\includegraphics[width=0.4\columnwidth]{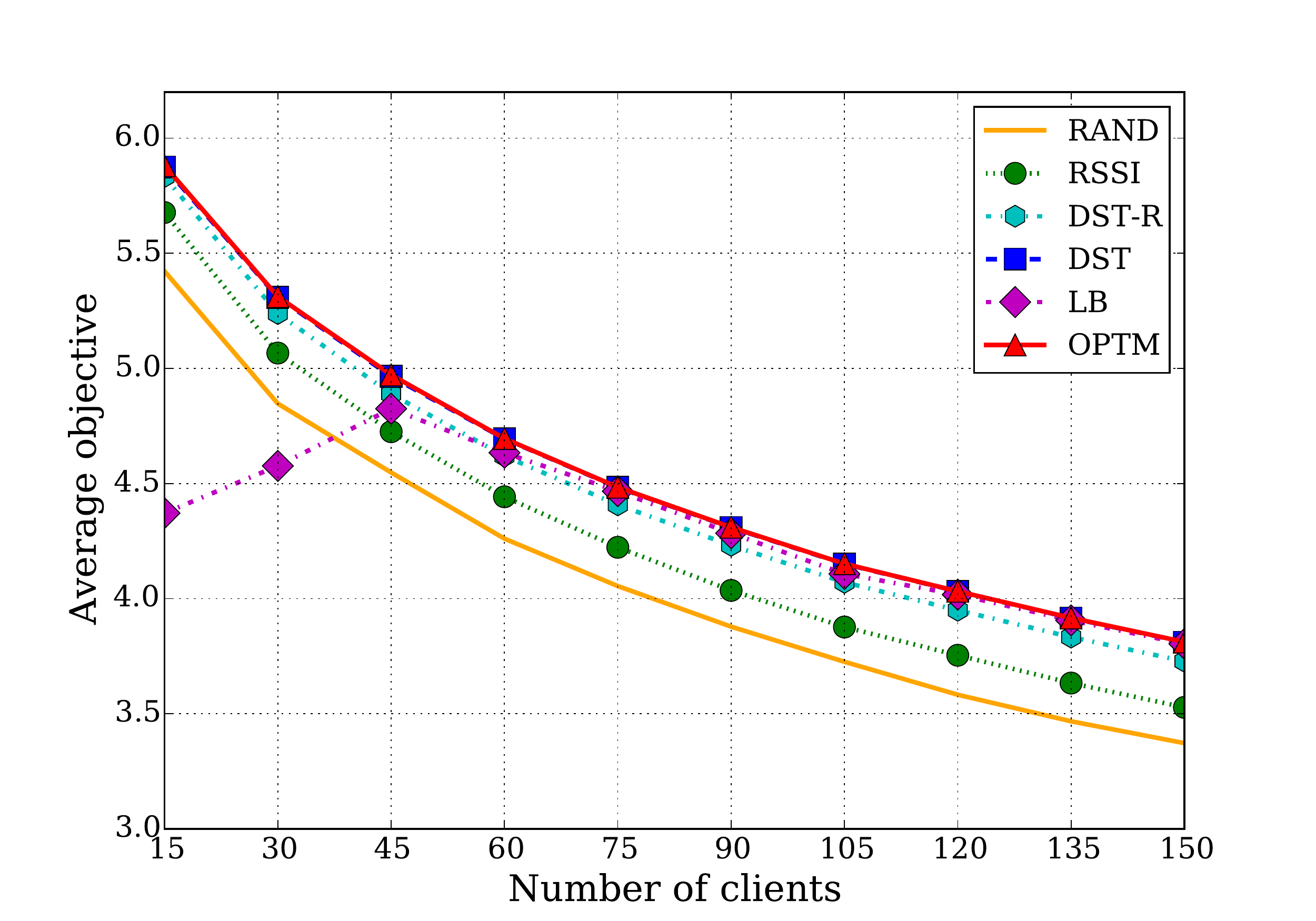}
\label{fig.obj-varyRAll}}
\subfloat[\footnotesize Half bottom clients and relays]{
\includegraphics[width=0.4\columnwidth]{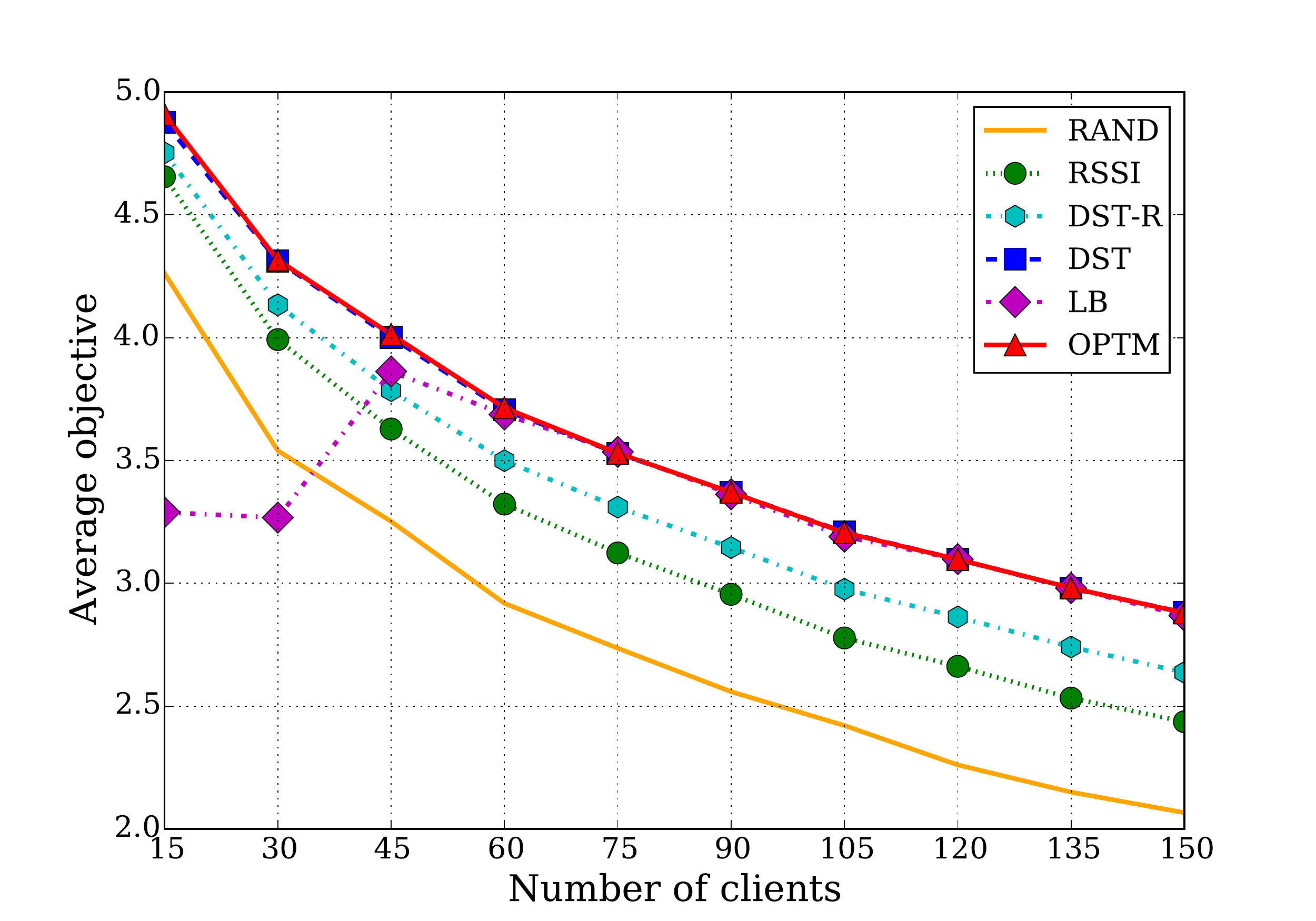}
\label{fig.obj-varyRHalf}
}
\caption{Average value of the objective function against the number of clients with $M/N=3$: (a) considers all the clients and relays, (b) considers half of the clients and relays with lower rates.}
\label{fig.obj-varyR}
\vspace{-0.8cm}
\end{figure}

Fig.~\ref{fig.obj-varyR} illustrates the average objective value, $\sum_{i\in\mathcal C} \sum_{k\in\mathcal A} \log (r_i y_{ik})/(M+N)$, obtained by the different association policies, when we fix the ratio of the numbers of clients and relays, namely $M/N = 3$. As shown in Fig.~\ref{fig.obj-varyR}, the average of the resulting objective value decreases with the number of clients. Consistent to Fig.~\ref{fig.rate-ratio}, as shown in Fig.~\ref{fig.obj-varyRHalf}, the proposed DST provides substantially larger objective than the DST without relaying for the clients and relays with lower rates. Again, the OPTM and DST almost overlap. Furthermore, the LB has the worst performance when the number of clients is less than 30, as shown in Fig.~\ref{fig.obj-varyR}.

\begin{figure}[t]
    \centering
    \includegraphics[width = 0.4\columnwidth]{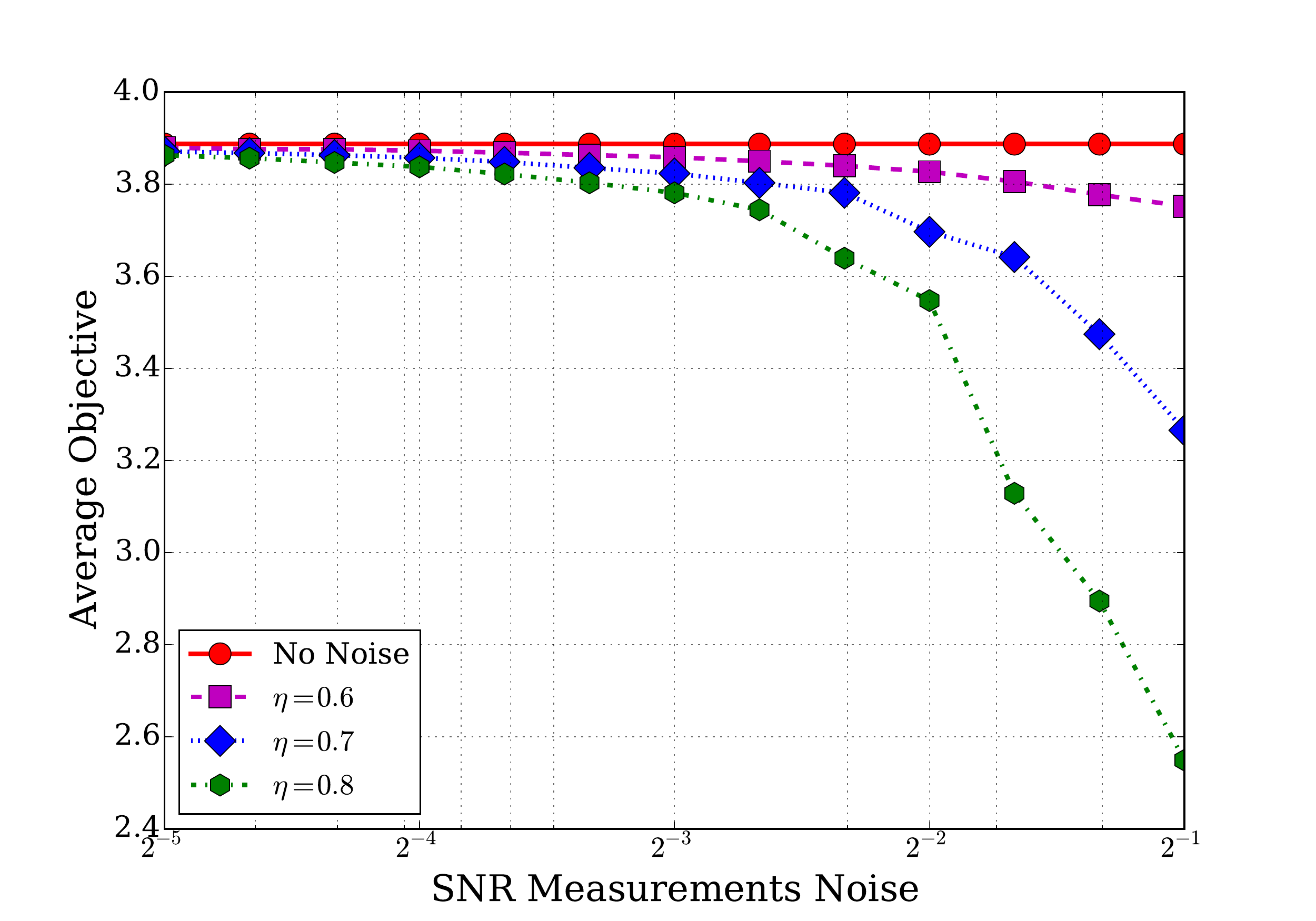}
    \caption{Illustration of the average objective value as a function of the variance of the SNR measurement noise $\sigma$.}
    \label{fig.var-perfm}
\end{figure}

To evaluate the impact of imperfect channel state information on the association performance, we assume that the measured $\widetilde{\textrm{SNR}}_{ij} = \textrm{SNR}_{ij} + e_{ij}$, where $e_{ij}$ is the error due to estimation and limited feedback channel. We further assume that $e_{ij}$ follows a zero-mean Gaussian distribution with variance $\sigma$. We denote $\sigma$ as SNR measurement noise. Fig.~\ref{fig.var-perfm} illustrates the average objective value as a function of the SNR measurement noise $\sigma$. Clearly, when $\sigma$ is large, constraints~\eqref{eq:const-stoch-linkcap-ij}$\sim$\eqref{eq:const-stoch-linkcap-j} force the clients to adopt a conservative data rate. Higher measurement noise variance increases the uncertainty on the final association and consequently increases the gap between the solution of \eqref{eq:stoch-optim-maxflow} and the optimal solution, once precise SNR values are available. However, for relatively high error in order of 0.1, the optimality gap is almost negligible.

\begin{figure}[t!]
\centering
\subfloat[]{
\includegraphics[width=0.4\columnwidth]{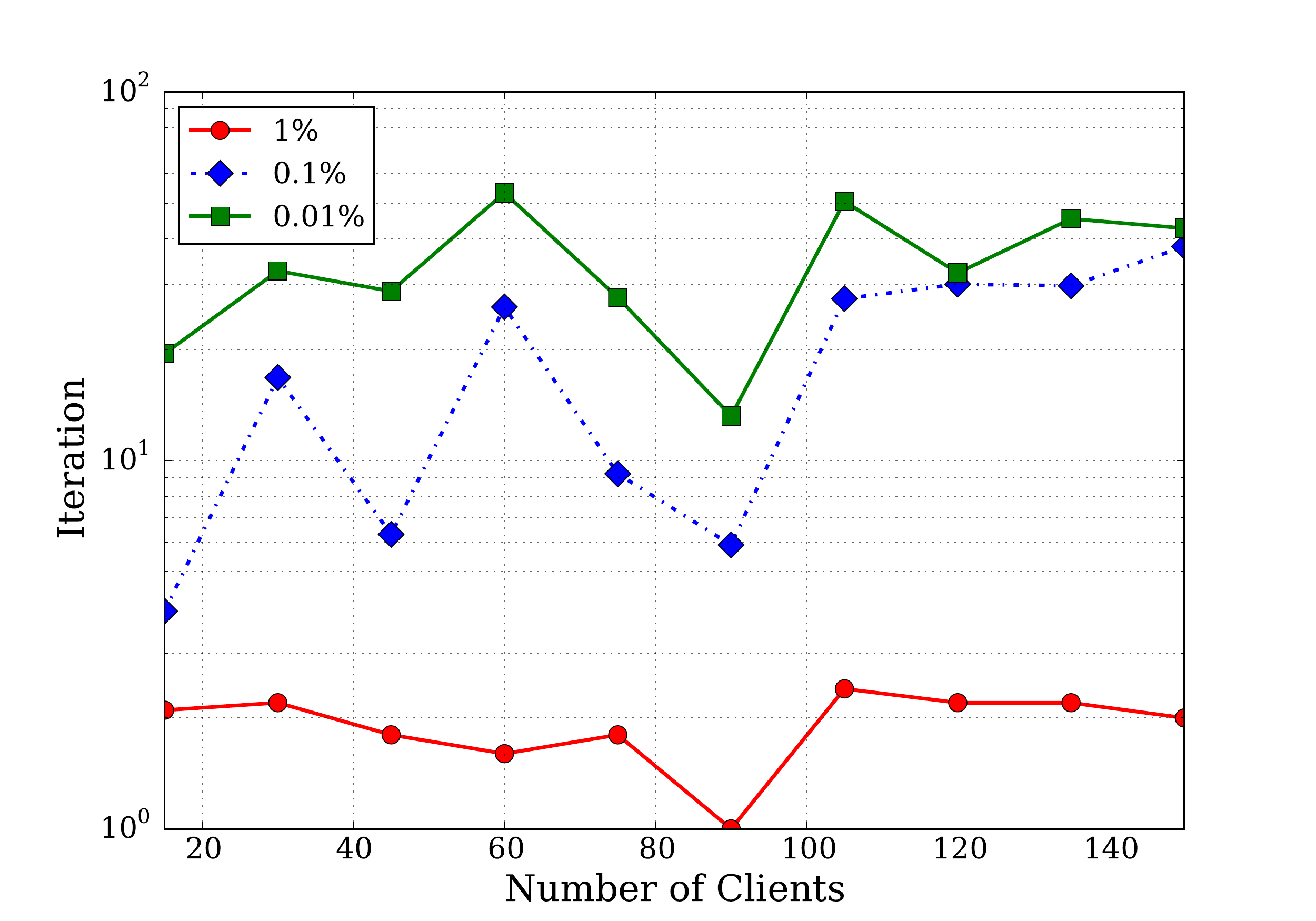}
\label{fig.iteration-varyC}}
\subfloat[]{
\includegraphics[width = 0.4\columnwidth]{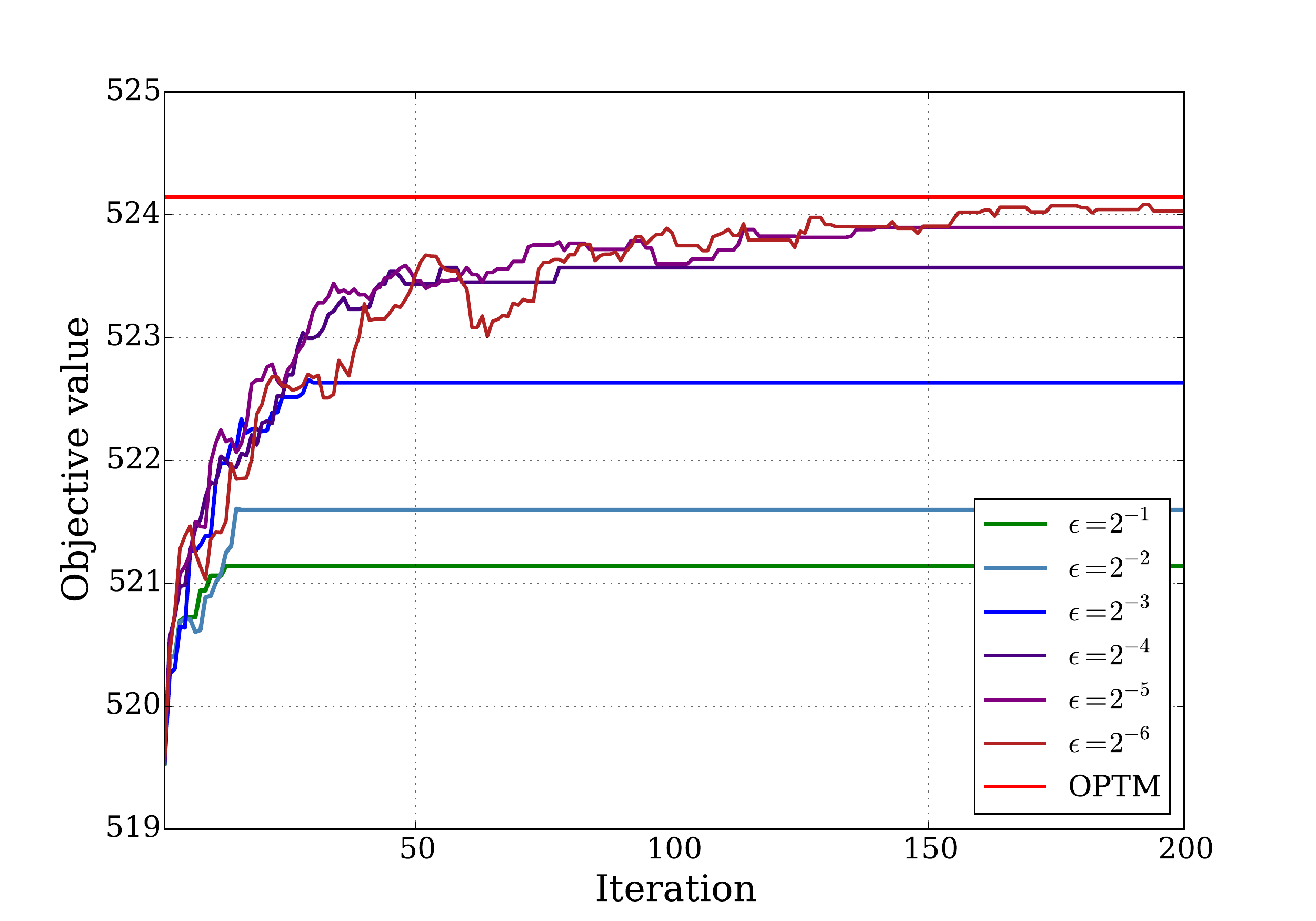}
\label{fig.obj-varyEpsilon}
}
\caption{(a) Illustration the average numbers of iterations required by Algorithm~\ref{alg.primal-dual} to converge to the optimal solution within 1\% (red circle), 0.1\% (blue diamond), and 0.01\% (green square) error bound varying the number of the clients, respectively. (b) Illustration of the performance of Algorithm~\ref{alg_d} varying the value of $\epsilon$.}
\end{figure}

\begin{figure}[t]
    \centering
    \includegraphics[width = 0.4\columnwidth]{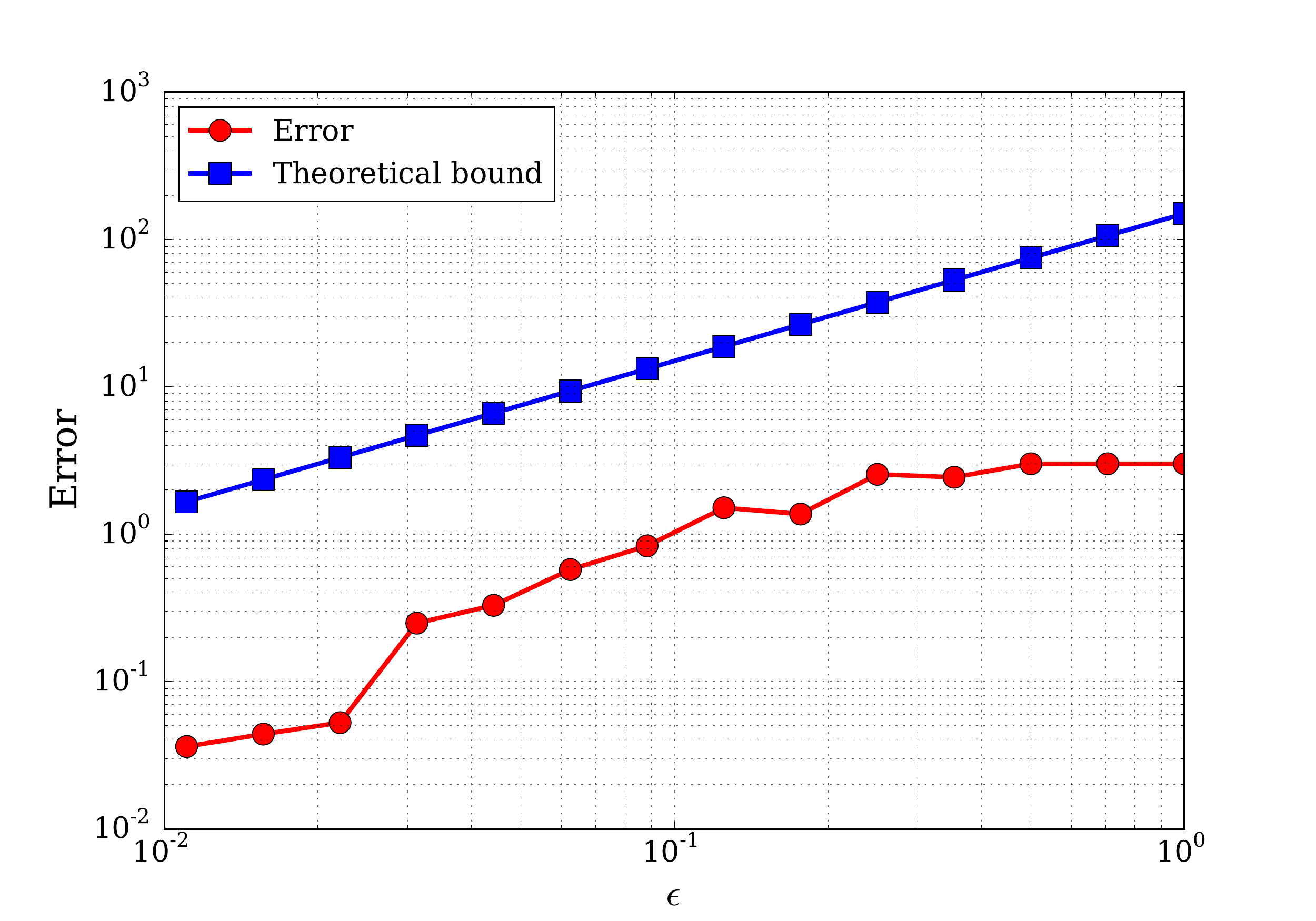}
    \caption{Illustration of the errors between the optimal objective values of optimization problem~\eqref{eq:true-prob} and that obtained by Algorithm~\ref{alg_d}, where the blue curve corresponds to the theoretical bound proposed by Proposition~\ref{prop.CS-Mepsilon}.}
    \label{fig.err-varyEpsilon}
    \vspace{-0.8cm}
\end{figure}

We show that by a proper choice of the desired convergence error, the number of iterations for Algorithms~\ref{alg.primal-dual} and \ref{alg_d} is small. Thus, these algorithms can be easily implemented on top of the beaconing mechanisms of existing mmWaves standards~\cite{802_11ad,802_15_3c}. Fig.~\ref{fig.iteration-varyC} shows the average number of iterations required by Algorithm~\ref{alg.primal-dual} in DST to converge to the optimal solution within 1\% error bound. From the figure, a moderate number of iterations is enough for Algorithm~\ref{alg.primal-dual} to achieve convergence. Fig.~\ref{fig.obj-varyEpsilon} illustrates the convergence performance of Algorithm~\ref{alg_d} in the DST by varying the value of $\epsilon$. When the value of $\epsilon$ decreases, the final solution becomes closer to the optimal solution at the expense of a lower convergence speed, as predicted by Proposition~\ref{prop.finity-convergence}: the auction algorithms are faster for larger $\epsilon$ values. To elaborate more, we report in Fig.~\ref{fig.err-varyEpsilon} the average error of the final solution of Algorithm~\ref{alg_d}. The results indicate that the error between the optimal objective values and that obtained by Algorithm~\ref{alg_d} raises with $\epsilon$. Moreover, the errors are always less than the upper bound provided in Proposition~\ref{prop.CS-Mepsilon}.

\vspace{-0.5cm}
\section{Conclusions}
\label{sec:conclusion}

In this paper, the problem of joint optimizing association of the clients to APs and relays and resource allocation in the APs in mmWave networks was investigated. The objective was to maximize the logarithmic utility of the rates for the clients in the network considering the load balancing in the APs. The resulting optimization problem is combinatorial and non-convex. We showed that it can be transformed into a {multi-dimensional} assignment problem. Then, a novel distributed algorithm based on the auction algorithm was developed to solve the problem. The performance of the proposed algorithm was investigated and illustrated in comparison to standard approaches through theoretical and numerical analysis. The results showed that the association with relaying can substantially improve the mmWave performance, that standardized methods are quite sub-optimal, and that, in general, relaying may play an important role. Our results indicate that the proposed solutions could be well applied in the forthcoming mmWave networks, envisioned to play a key role in future wireless systems.

\vspace{-0.5cm}
\bibliographystyle{IEEEtran}
\bibliography{ref}

\end{document}